\documentclass[10pt]{article}
\usepackage[colorinlistoftodos]{todonotes}
\usepackage{graphicx}
\usepackage[T1]{fontenc}
\usepackage[utf8]{inputenc}
\usepackage{amssymb}
\usepackage{amsfonts}
\usepackage{dsfont}
\usepackage{mathtools}
\usepackage{amsthm}
\usepackage{amsmath}
\usepackage{relsize}
\usepackage{textcomp}
\usepackage{eurosym}
\usepackage{stmaryrd}
\usepackage{xcolor}
\usepackage[multiple]{footmisc}
\usepackage{pdflscape}
\usepackage[title]{appendix}

\usepackage[font=small,labelfont=bf]{caption}

\usepackage{bigints}
\usepackage{geometry}
\begin{document}

\title{Price-Aware Automated Market Makers: Models Beyond Brownian Prices and Static Liquidity}

\author{Philippe \textsc{Bergault}\footnote{Université Paris Dauphine-PSL, Ceremade, Place du Maréchal de Lattre de Tassigny, 75116 Paris, France, bergault@ceremade.dauphine.fr.} \and Louis \textsc{Bertucci}\footnote{Institut Louis Bachelier, Palais Brongniart, 28 Place de la Bourse, 75002 Paris, France, louis.bertucci@institutlouisbachelier.org.} \and David \textsc{Bouba}\footnote{Swaap Labs, d@swaap.finance.} \and Olivier \textsc{Guéant}\footnote{Université Paris 1 Panthéon-Sorbonne, Centre d'Economie de la Sorbonne, 106 Boulevard de l'Hôpital, 75642 Paris Cedex~13, France, olivier.gueant@univ-paris1.fr.} \and Julien \textsc{Guilbert}\footnote{Swaap Labs, julien@swaap.finance.}}

\date{}

\maketitle
\setlength\parindent{0pt}

\begin{abstract}

In this paper, we introduce a suite of models for price-aware automated market making platforms willing to optimize their quotes. These models incorporate advanced price dynamics, including stochastic volatility, jumps, and microstructural price models based on Hawkes processes. Additionally, we address the variability in demand from liquidity takers through models that employ either Hawkes or Markov-modulated Poisson processes. Each model is analyzed with particular emphasis placed on the complexity of the numerical methods required to compute optimal quotes.\\

\vspace{5mm}

\end{abstract}

\section{Introduction}

Market makers, including both human agents and algorithms, are key participants in financial markets standing ready to buy and sell securities or currency pairs. They are crucial in facilitating transactions by bridging the gap between buyers and sellers, whose requests or orders may not coincide due to asynchronous submissions.\\

The challenges faced by market makers, and the broader microeconomics of market making, have been rigorously examined since the 1980s from two perspectives. The first is information asymmetries, initially explored by Copeland and Galai \cite{copeland1983information} and subsequently by Glosten and Milgrom \cite{glosten1985bid}. The second perspective focuses on inventory management, as initially investigated by Ho and Stoll \cite{ho1981optimal,ho1983dynamics}.\\

In the literature concerning information asymmetries, the bid-ask spread is primarily viewed as an informational phenomenon. Faced with the risk of adverse selection by informed traders, market makers -- even those who are risk-neutral -- must maintain a bid-ask spread to sustain their viability. This body of literature often employs metaphorical (toy) models to elucidate the fundamental informational factors that drive bid-ask spreads.\\

When holding inventory, market makers are confronted with risks arising from uncertainties about future prices and the trading willingness of other market participants. Within this context, the bid-ask spread is utilized as a premium to compensate for the risks incurred while holding inventory. Quantitative models in the inventory management literature are often designed to maximize the expected profit and loss (PnL) of market makers, while also managing inventory levels to remain as close to zero as possible, primarily through the strategic setting of their quotes. The body of literature on inventory management has seen significant advancements since Avellaneda and Stoikov's 2008 article \cite{avellaneda2008high}, which reinvigorated the modeling framework originally introduced by Ho and Stoll nearly three decades earlier. Within the quantitative finance community, researchers have built upon the foundational elements of Ho and Stoll’s seminal works, enriching models with increasingly sophisticated features over the years. These enhancements include the integration of alpha signals, varying order sizes, client tiering strategies, the internalization-versus-externalization dilemma, consideration of market impact, introduction of uncertainty in parameters, and attempts to address the challenges of adverse selection (see \cite{barzykin2023algorithmic, barzykin2021market, bergault2021size, cartea2017algorithmic, cartea2015algorithmic, cartea2014buy, gueant2016financial, gueant2013dealing}). Furthermore, there has been an effort to extend these models to multi-asset settings, alongside the development of sophisticated numerical and approximation techniques (see \cite{barzykin2022dealing, bergault2018closed, bergault2021size, gueant2017optimal, gueant2019deep}).\\

These market making models have shown effectiveness in over-the-counter (OTC) markets, such as those for corporate bonds and foreign exchange. With the recent rise of decentralized finance (DeFi) and the emergence of Automated Market Makers (AMMs) in cryptocurrency markets -- which provide alternatives to centralized exchange platforms based on limit order books -- it appears promising that these models could be adapted to optimize price-aware AMMs.\footnote{Such adaptations must take into account specific characteristics like the use of the buy-and-hold (Hodl) benchmark and the requirement for inventories to remain nonnegative.} Price-aware AMMs mark a new generation in the DeFi arena. Unlike the classical Constant Function Market Makers (CFMMs), which leave price discovery to liquidity takers, including arbitrageurs, and stay passive in the price formation process, price-aware AMMs utilize current market estimates to inform their pricing strategies. A leading study in this new direction is \cite{bergault2022automated}, which also coined the concept of efficient market making strategies. In \cite{bergault2022automated}, however, the model assumes log-normal price dynamics and static liquidity, which do not optimally represent the characteristics of many cryptocurrency pairs. This paper aims therefore to exhibit a suite of models extending that introduced in~\cite{bergault2022automated} by incorporating more complex price dynamics and dynamic liquidity.\\

In the literature on market making, asset prices are commonly modeled using Brownian motions or geometric Brownian motions, sometimes with a drift or a stochastic alpha component. There are exceptions, however. For instance, \cite{faycal} examines a multidimensional Ornstein-Uhlenbeck dynamics suitable for cointegrated assets, and \cite{rosenbaum2022multi} attempts to integrate stochastic volatility into market making models. Our paper explores the effects of employing models with price jumps and stochastic volatility on market making equations. Our primary focus is on dynamics from the option pricing literature, particularly the Heston-Bates model (see \cite{bates1996jumps}) and the Stein-Stein model with jumps (see \cite{stein1991stock}). We demonstrate that while these models necessitate the addition of a state variable, they do not preclude the use of the quadratic Hamiltonian approximation technique introduced in \cite{bergault2018closed} to approximate value functions. However, the resulting Riccati equations are replaced by nonlinear parabolic partial differential equations. We also show that the quadratic Hamiltonian approximation technique can be applied to several price models based on Hawkes processes recently introduced in the market microstructure literature.\\

Besides price dynamics, we address the impact of stochastic liquidity, diverging from the common assumption in most market making models of a constant flow of liquidity takers for fixed bid-ask spread and skew. Given the stochastic nature of liquidity and the observed autocorrelation in trade signs, Hawkes processes represent a natural choice beyond traditional Poisson processes (see, for instance, \cite{jusselin}). However, employing Hawkes processes presents a challenge as it conflicts with the quadratic Hamiltonian approximation technique, which reduces computational complexity. As an alternative, we propose Markov-modulated Poisson processes (MMPPs), as explored in the recent paper \cite{bergault2023modeling}.\\

In this paper, we present a comprehensive suite of models for automated market makers. Our primary focus is on detailing the Hamilton-Jacobi-Bellman equations that characterize the value function of price-aware automated market makers and the associated optimal quotes, exploring different models for price and liquidity dynamics. Section 2 introduces the general modeling framework that underpins all the models we discuss. Then, Section 3 explores various models for price dynamics, while Section 4 examines two ways to model stochastic liquidity.

\section{General modelling framework}
\label{secgeneral}

In this paper, we consider a filtered probability space $\left( \Omega, \mathcal{F},\mathbb{P}; \mathbb{F}= (\mathcal{F}_{t})_{t\geq 0} \right)$ satisfying the usual conditions. All the random variables and stochastic processes are defined on this probability space.\\

We consider a two-currency automated market maker which offers quotes to liquidity takers (LTs) based on a reference price\footnote{This reference price may be derived from an off-chain market feed or from an on-chain price oracle, depending on the setup.} and markups\footnote{The markups are typically computed offchain whenever the computational complexity becomes high. The underlying algorithm can be public or private depending on the setup.}. Throughout this paper, the two currencies will be referred to as currency~$0$ and currency~$1$, with currency~$0$ serving as the accounting currency. The process $(S_t)_{t\in \mathbb{R}_+}$ denotes the reference price, expressing the value of currency~$1$ in terms of currency~$0$.\\

Our goal in all the models presented below is to develop efficient markups for liquidity providers (LPs) as defined in~\cite{bergault2022automated}. Specifically, we seek to find markups that maximize the expected excess Profit and Loss (PnL) of LPs relative to the Hodl benchmark,\footnote{The excess PnL relative to Hodl represents the PnL of a representative liquidity provider between two liquidity provision or redemption events, compared to the PnL that would have been realized had the LP remained outside the pool. Other benchmarks and decompositions of LPs' PnL have been proposed in the literature. For CFMMs, a notable decomposition is presented in \cite{cartea2023predictable}, introducing the concept of predictable loss. In a specific model, \cite{milionis2022automated} introduced the related concept of loss-versus-rebalancing and regarded it as a benchmark, albeit an extreme one due to its assumption of back-to-back trading. In the context of price-aware AMMs, the relevant benchmark remains Hodl. However, the idea of reducing risk by enabling AMMs to trade on other platforms is highly relevant. This strategy is reminiscent of the classical internalization versus externalization trade-off in traditional market making (see \cite{barzykin2022dealing, barzykin2021market}).} while ensuring that the variance of this excess PnL remains below a predetermined level.\\

Regarding transactions, we consider the same kind of models as in \cite{bergault2022automated}. Transaction sizes are labelled in the accounting currency and transaction prices are decomposed into two parts: one part corresponding to the reference exchange rate and another part corresponding to a markup (that might very rarely be a discount) that is accounted in currency $0$, whatever the side of the transaction, for the sake of simplicity. More precisely, if a liquidity taker wants to sell $z$ coins of currency $0$ at time $t$, then $z/S_t$ coins of currency $1$ will be offered to them and $z\delta^{0,1}(t,z)$ extra coins of currency $0$ will be asked as a markup. Symmetrically, if a liquidity taker wants to buy $z$ coins of currency $0$ at time $t$, then $z/S_t$ coins of currency $1$ will be asked and $z\delta^{1,0}(t,z)$ out of the total of $z$ coins of currency $0$ will not be transferred to them.\\

We assume that the markups $\left(\delta^{0,1}, \delta^{1,0}\right)$ belong to
\begin{equation}
\begin{split}
\mathcal{A}:= \Bigg\lbrace  \left(\delta^{0,1},\delta^{1,0}\right) : \Omega \times [0,T]& \times \mathbb{R}_{+}^{*} \mapsto \mathbb{R}^{2} \bigg| \left(\delta^{0,1},\delta^{1,0}\right)  \text{ is } \mathcal{P} \otimes \mathcal{B}(\mathbb{R}_{+}^{*})\text{-measurable }\Bigg\rbrace, \nonumber
\end{split}
\end{equation}
where $\mathcal{P}$ denotes the $\sigma$-algebra of $\mathbb{F}$-predictable subsets of $\Omega \times[0,T]$ and $\mathcal{B}(\mathbb{R}_{+}^{*})$ denotes the Borelian sets of $\mathbb{R}_{+}^{*}$.\\

To simplify the analysis, we accumulate these markups in a process $(X_t)_{t \in [0,T]}$, separated from the pool reserves, whose dynamics is
$$dX_t =  \int_{z \in \mathbb R_+^*} z\delta^{0,1} (t,z) J^{0,1}(dt,dz) + \int_{z \in \mathbb R_+^*} z\delta^{1,0} (t,z) J^{1,0}(dt,dz),$$
with $X_0=0$, where $J^{0,1}(dt,dz)$ and $J^{1,0}(dt,dz)$ are two $\mathbb R_+^*$-marked point processes modelling respectively transactions through which the AMM sells currency $1$ and receives currency $0$ (for $J^{0,1}(dt,dz)$) and transactions through which the AMM sells currency $0$ and receives currency $1$ (for $J^{1,0}(dt,dz)$).\\

The dynamics of the reserves are then given by:
$$dq^0_t =   \int_{z \in \mathbb R_+^*} z \left(J^{0,1}(dt,dz) -  J^{1,0}(dt,dz) \right) \quad \text{and} \quad dq^1_t =  \int_{z \in \mathbb R_+^*} \frac z{S_t} \left(J^{1,0}(dt,dz) -  J^{0,1}(dt,dz) \right).$$

We assume that the processes $J^{0,1}(dt,dz)$ and $J^{1,0}(dt,dz)$ have known intensity kernels given respectively by $(\zeta^{0,1}_t(dz))_{t\in \mathbb R_+}$ and $(\zeta^{1,0}_t(dz))_{t\in \mathbb R_+}$, verifying
$$\zeta^{0,1}_t(dz) = \Lambda^{0,1}\left(t,z, \delta^{0,1} (t,z)\right)\mathds{1}_{\{q^1_{t-}\ge \frac{z}{S_t}\}}m(dz) \quad \text{and} \quad \zeta^{1,0}_t(dz) = \Lambda^{1,0}\left(t,z, \delta^{1,0} (t,z)\right)\mathds{1}_{\{q^0_{t-}\ge z\}}m(dz),$$
where:
\begin{itemize}
\item $\Lambda^{0,1} : (t,z,\delta) \in \mathbb R_+ \times \mathbb R^*_+ \times \mathbb R \mapsto \mathbb R_+^*$ and $\Lambda^{1,0} : (t,z,\delta) \in \mathbb R_+ \times \mathbb R^*_+ \times \mathbb R \mapsto \mathbb R_+^*$ are intensity functions decreasing in $\delta$ and satisfying the classical technical assumptions of intensity functions in the market making literature (see \cite{bergault2021size});
\item the indicator functions represent the impossibility for the AMM to propose exchange rates for transactions that cannot occur because reserves are too low in the demanded currency;
\item $m$ is a measure on $\mathbb R_+$ (typically Lebesgue or discrete).
\end{itemize}
In the standard literature on OTC market making (see \cite{barzykin2021market} or \cite{bergault2021size}, for instance), these intensity functions (which correspond to the demand curve of LTs) are assumed of the logistic type. In our general framework where liquidity may not be constant, we consider
$$\Lambda^{0,1}(t,z,\delta)=\lambda^{0,1}_t \phi^{0,1}(z) f^{0,1}(z,\delta) \quad \text{and} \quad \Lambda^{1,0}(t,z,\delta)=\lambda^{1,0}_t \phi^{1,0}(z) f^{1,0}(z,\delta)$$
where:
\begin{itemize}
\item $\lambda^{0,1}_t$ and $\lambda^{1,0}_t$ describe the height of the demand curve;
\item $\phi^{0,1}(z)$ and $\phi^{1,0}(z)$ are the probability density functions (with respect to $m$) of the size of trade inquiries;
\item $f^{0,1}: (z,\delta) \in \mathbb R^*_+ \times \mathbb R \mapsto  \frac{1}{1+e^{a^{0,1}(z) + b^{0,1}(z) \delta}}$ and $f^{1,0} : (z,\delta) \in \mathbb R^*_+ \times \mathbb R \mapsto \frac{1}{1+e^{a^{1,0}(z) + b^{1,0}(z) \delta}}$ describe the logistic shape of the demand curve, in particular the sensitivity to the markups which may depend on sizes. 
\end{itemize}

If we mark the reserve of currency $1$ with the reference exchange rate, the total value associated with the markups and the reserves in the pool at time~$T$ is given by $X_T + q^0_T + q^1_T S_T$. To evaluate the PnL of a liquidity provider compared to that of an agent who would have held the coins outside of the AMM (Hodl benchmark), we need to subtract the current value of the initial reserves, i.e., $q^0_0 + q^1_0 S_T$. Therefore, the excess PnL with respect to Hodl at time $T$ is expressed as:
\begin{align}
\Pi_T &= X_T + \left(q^0_T - q^0_0\right)  + \left(q^1_T - q^1_0\right) S_T\nonumber\\
&= \int_0^T \int_{z \in \mathbb R_+^*} z\delta^{0,1} (t,z) J^{0,1}(dt,dz) + \int_0^T\int_{z \in \mathbb R_+^*} z\delta^{1,0} (t,z) J^{1,0}(dt,dz)\nonumber\\
&\quad + \int_0^T \int_{z \in \mathbb R_+^*}\!\! z \left(J^{0,1}(dt,dz) -  J^{1,0}(dt,dz) \right) +  \int_0^T (q^1_t-q^1_0) dS_t \nonumber\\
&\quad +  \int_0^T\int_{z \in \mathbb R_+^*}\!\!  z \left(J^{1,0}(dt,dz) -  J^{0,1}(dt,dz) \right)\nonumber\\
&= \int_0^T \int_{z \in \mathbb R_+^*} z\delta^{0,1} (t,z) J^{0,1}(dt,dz) + \int_0^T\int_{z \in \mathbb R_+^*} z\delta^{1,0} (t,z) J^{1,0}(dt,dz) +  \int_0^T (q^1_t-q^1_0) dS_t.\label{PnL}
\end{align}

Given a time horizon $T$, we can define efficient market making strategies as those maximising the expected value of the above excess PnL, subject to an upper bound on its variance. These strategies correspond to those maximizing expressions of the form
$$\mathbb{E}[\Pi_T] - \frac{\gamma}{2} \mathbb{V}[\Pi_T],$$ for $\gamma$ spanning $\mathbb{R}_+$. Traditional optimization problems of this form are known to suffer from time inconsistency due to the variance term. To address this issue, we adopt very close but alternative objective functions that penalize risk. In Eq.~\eqref{PnL}, risk originates from two main sources: the stochastic nature of trade arrivals and price fluctuations. Existing literature on market making (such as \cite{gueant2017optimal} or more recently \cite{barzykin2021market}) suggests that price risk is the dominant factor. Consequently, our subsequent analysis focuses on mitigating price risk. To work with a time-consistent objective function, we replace the variance of $\Pi_T$ in the original optimization problems with the quadratic variation of $\int_0^T (q^1_t-q^1_0) dS_t$, i.e., $\int\limits_{0}^{T} (q_t^1 - q_0^1)^2 d\langle S_t \rangle$. This modification leads to the following objective functions (for $\gamma \ge 0$):
\begin{align*}
\mathbb{E}\Bigg[&\int\limits_{0}^{T}\Bigg\lbrace \int_{z \in \mathbb R_+^*} \Big(z\delta^{0,1}(t,z)  \Lambda^{0,1}(t,z,\delta^{0,1}(t,z))\mathds{1}_{\{q^1_{t-}\ge \frac{z}{S_t}\}}+z\delta^{1,0}(t,z)  \Lambda^{1,0}(t,z,\delta^{1,0}(t,z))\mathds{1}_{\{q^0_{t-}\ge z\}}  \Big)m(dz)\Bigg\rbrace dt\nonumber\\
& + \int\limits_{0}^{T} (q_t^1 - q_0^1)dS_t - \frac \gamma 2 \int\limits_{0}^{T} (q_t^1 - q_0^1)^2 d\langle S_t \rangle  \Bigg].
\end{align*}

In the presence of the penalty term $\frac{\gamma}{2} \int\limits_{0}^{T} (q_t^1 - q_0^1)^2 d\langle S_t \rangle$ -- at least for $\gamma > 0$ -- the market making strategy is naturally inclined to adjust prices such that the reserves remain close to their initial values. Therefore, it is justified, following the approach by \cite{bergault2022automated}, to simplify the problem by omitting the indicator functions.\footnote{In practical applications, requests that would deplete the reserves on one side of the pool can simply be prohibited.} This removal of indicator functions also reduces the dimensionality of the problem, as it decreases the number of state variables required. The stochastic optimal control problems are then formulated (for $\gamma \ge 0$) as follows:
\begin{align}
\sup_{\left(\delta^{0,1},\delta^{1,0}\right) \in \mathcal A}&\mathbb{E}\Bigg[\int\limits_{0}^{T}\Bigg\lbrace \int_{z \in \mathbb R_+^*} \Big(z\delta^{0,1}(t,z)  \Lambda^{0,1}(t,z,\delta^{0,1}(t,z))+z\delta^{1,0}(t,z)  \Lambda^{1,0}(t,z,\delta^{1,0}(t,z))  \Big)m(dz)\Bigg\rbrace dt\nonumber\\
& \quad + \int\limits_{0}^{T} (q_t^1 - q_0^1)dS_t - \frac \gamma 2 \int\limits_{0}^{T} (q_t^1 - q_0^1)^2 d\langle S_t \rangle  \Bigg].\label{sto_opt_cont}
\end{align}

The nature and dimensionality of the stochastic optimal control problems \eqref{sto_opt_cont} are influenced by two types of assumptions: (i) assumptions regarding the dynamics of $(S_t)_t$, and (ii) assumptions concerning the dynamics of liquidity, specifically $(\lambda^{0,1}_t)_t$ and $(\lambda^{1,0}_t)_t$. Section 3 is dedicated to discussing models for the dynamics of~$(S_t)_t$. In Section 4, we explore dynamic models for liquidity.

\section{Efficient market making strategies with advanced price models}
\label{sec_prices}

\subsection{Introduction}

The recent paper \cite{bergault2022automated} introduces efficient price-aware market making strategies within a framework assuming constant liquidity and log-normal price distributions. In this section, we retain the assumption of constant liquidity, meaning that $\lambda_t^{0,1} = \lambda^{0,1}$ and $\lambda_t^{1,0} = \lambda^{1,0}$ are held constant,\footnote{Throughout Section \ref{sec_prices}, we omit the time variable in the intensity functions $\Lambda^{0,1}$ and $\Lambda^{1,0}$.} but we broaden our analysis to include models beyond log-normal price distributions. To effectively integrate a dynamic price model into our optimal control framework and avoid the curse of dimensionality, it is crucial to employ continuous-time price models that are both Markovian and low-dimensional.\\

In the field of derivatives pricing, the Heston model is among the most well-known stochastic volatility models. It employs a Cox-Ingersoll-Ross (CIR) process to model the square of the volatility. This model has been further extended to incorporate jumps in price dynamics, resulting in what is commonly referred to as the Heston-Bates model. In Section~\ref{subHB}, we explore the stochastic optimal control problems \eqref{sto_opt_cont} within the context of Heston-Bates price dynamics. Another seminal model in the option pricing literature is the Stein-Stein model, which uses an Ornstein-Uhlenbeck process to model volatility. Section~\ref{subSSJ} examines the stochastic optimal control problems \eqref{sto_opt_cont} within the Stein-Stein framework, specifically in its Schöbel and Zhu extension \cite{schobel1999stochastic}, to which we have added jumps. In addition to models from the derivatives pricing literature, continuous-time models have also been proposed in the market microstructure literature. Section~\ref{subH} examines the stochastic optimal control problems \eqref{sto_opt_cont} within the framework of a Hawkes model for price dynamics. Section~\ref{subZH} proposes an extension with Z-Hawkes processes following a microstructural model proposed in \cite{ZH}.\\

\subsection{Introducing stochastic volatility and jumps through the Heston-Bates model}\label{subHB}

\subsubsection{Model, Hamilton-Jacobi-Bellman equation and optimal markups}

In this section, we consider for the price process $(S_t)_{t\in \mathbb R_+}$ the following model:
$$
\begin{cases}
    dS_t = \mu S_t dt + \sqrt{\nu_t} S_t dW^S_t + S_t\int_{\eta \in \mathbb R} \eta \bar J(dt,d\eta)\\
    d\nu_t = k (\bar \nu - \nu_t) dt + \sqrt{\nu_t} \xi dW^\nu_t
\end{cases}
$$
where:
\begin{itemize}
\item $\mu \in \mathbb R$ is a known deterministic drift;
\item $(\nu_t)_{t\in \mathbb R_+}$ is the instantaneous variance process with values in~$\mathbb R_+$;
\item $\bar J$ is a marked point process with constant-in-time kernel $\bar \lambda \zeta(d\eta)$ where $\bar \lambda$ is the rate of jumps and $\zeta$ the probability measure of jump sizes in relative value (assumed to be such that $\int_{\eta \in \mathbb R} \eta \zeta(d\eta) = 0$);
\item $k>0$ is the rate of mean reversion of the instantaneous variance;
\item $\xi>0$ measures the stochasticity of the instantaneous variance process;
\item $\bar \nu >0$ is the long-term instantaneous variance;
\item $\left(W^S_t, W^\nu_t \right)_{t\in \mathbb R_+}$ is a 2-dimensional standard Brownian motion independent of $\bar J$ with $d\langle W^S, W^\nu \rangle_t = \rho dt$ for $\rho \in [-1,1]$.\\
\end{itemize}

In this framework, it is interesting to introduce $Y_t = (q_t^1 - q_0^1) S_t$. We have indeed that $(Y_t, \nu_t)_{t \in \mathbb R_+}$ is Markovian since
$$ 
dY_t = \mu Y_t dt + \sqrt{\nu_t} Y_t dW^S_t + Y_t \int_{\eta \in \mathbb R} \eta \bar J(dt,d\eta)+  \int_{z \in \mathbb R_+^*} z \left(J^{1,0}(dt,dz) -  J^{0,1}(dt,dz) \right)$$ and \eqref{sto_opt_cont} writes
\begin{align}
\sup_{\left(\delta^{0,1},\delta^{1,0}\right) \in \mathcal A}&\mathbb{E}\Bigg[\int\limits_{0}^{T}\Bigg\lbrace \int_{z \in \mathbb R_+^*} \Big(z\delta^{0,1}(t,z)  \Lambda^{0,1}(z,\delta^{0,1}(t,z))+z\delta^{1,0}(t,z)  \Lambda^{1,0}(z,\delta^{1,0}(t,z))  \Big)m(dz)\Bigg\rbrace dt\nonumber\\
& + \int\limits_{0}^{T} \mu Y_t dt  - \frac{\gamma}{2} \int\limits_{0}^{T}\left(\nu_t + \bar \lambda {\bar \eta}^2 \right) Y_t^2 dt  \Bigg], \label{soc_HB}
\end{align}
where $\bar \eta = \left(\int_{\eta \in \mathbb R} \eta^2 \zeta(d\eta)\right)^{\frac 12}$.\\

For a given $\gamma$, we associate with \eqref{soc_HB} a value function $\theta:[0,T]\times \mathbb R \times \mathbb R_+ \rightarrow \mathbb{R}$ and the following Hamilton-Jacobi-Bellman equation:
\begin{equation}
\begin{cases}
 \!&0 = \partial_t \theta(t,y, \nu) + \mu y \left( 1 + \partial_y \theta(t,y, \nu) \right) - \frac{\gamma}{2} \left(\nu + \bar \lambda {\bar \eta}^2\right) y^2 + \frac 12 \nu y^2 \partial^2_{yy} \theta(t,y, \nu)\\
 \!& \qquad + k(\bar \nu - \nu)\partial_\nu \theta(t,y, \nu) +  \frac 12 \nu \xi^2 \partial^2_{\nu \nu} \theta(t,y, \nu) + \rho \nu y \xi \partial^2_{y\nu} \theta(t,y, \nu) \\
 \!& \qquad + \bar \lambda \int_{\eta \in \mathbb R}(\theta(t,y(1+\eta), \nu) - \theta(t,y, \nu)) \zeta(d\eta) \\
\!& \qquad + \text{\scalebox{0.6}[1]{$\bigint$}}_{\!\!\mathbb{R}_{+}^{*}} \left(z\lambda^{0,1}\phi^{0,1}(z)H^{0,1} \left(z,\frac{\theta(t,y,\nu) -  \theta(t,y- z, \nu) }{z}\right) + z\lambda^{1,0}\phi^{1,0}(z)H^{1,0} \left(z,\frac{\theta(t,y,\nu) -  \theta(t,y+z,\nu) }{z}\right) \right)m(dz),\\
\!&\theta(T,y,\nu) = 0,
\end{cases}
\label{eqn:HJBHB}
\end{equation}
where
$$
H^{0,1}:(z,p)\in\mathbb R_+^* \times \mathbb{R} \mapsto \ \underset{\delta}{\sup} f^{0,1}(z,\delta)(\delta - p) = \underset{\delta}{\sup} \frac{1}{1+e^{a^{0,1}(z) + b^{0,1}(z) \delta}}(\delta-p)$$
and
$$H^{1,0}:(z,p)\in\mathbb R_+^* \times \mathbb{R} \mapsto  \underset{\delta}{\sup} f^{1,0}(z,\delta)(\delta - p) = \underset{\delta}{\sup}\  \frac{1}{1+e^{a^{1,0}(z) + b^{1,0}(z) \delta}}(\delta-p).$$

Using the same ideas as in \cite{gueant2017optimal}, we see that for $i\neq j \in \{0,1\}$, the supremum in the definition of $H^{i,j}(z,p)$ is reached at a unique $\bar \delta^{i,j}(z,p)$ given by $$\bar \delta^{i,j}(z,p) = (f^{i,j})^{-1} \left(z, -\partial_p{H^{i,j}} (z,p)  \right)$$
where for all $z$, $(f^{i,j})^{-1}(z, .)$ denotes the inverse function of $f^{i,j}(z,.)$. Moreover, the markups that maximize our modified objective function are obtained in the following form
\begin{align}\label{optquotes01H}
 \delta^{0,1*}(t,z) = \bar \delta^{0,1} \left(z, \frac{\theta(t,Y_{t-}, \nu_t) -  \theta(t,Y_{t-}-z, \nu_t) }{z} \right)
 \end{align}
 and
\begin{align}\label{optquotes10H}
\delta^{1,0*}(t,z) = \bar \delta^{1,0} \left(z, \frac{\theta(t,Y_{t-}, \nu_t) -  \theta(t,Y_{t-}+z, \nu_t) }{z} \right).   
 \end{align}

\subsubsection{Discussion and numerical methods}

Under Heston-Bates dynamics for the price, computing the optimal markups primarily involves solving the partial integro-differential equation \eqref{eqn:HJBHB} to derive the value function $\theta$, which depends on time and two state variables: the spread to Hodl $y$ and the instantaneous variance $\nu$. Then, one can substitute the current value of $y$ and an estimate of the current value of $\nu$ (which is not directly observable) into Eqs. \eqref{optquotes01H} and \eqref{optquotes10H} to determine the optimal markups.\\

However, the value function $\theta$ cannot be determined in closed form and must instead be approximated numerically. This approximation can be achieved using a finite difference scheme on a rectangular grid, employing classical numerical methods such as operator splitting, discrete approximation of the integral terms, and an implicit scheme for the differential terms. It is therefore feasible to base a market making strategy on the numerical resolution of Eq. \eqref{eqn:HJBHB}. However, when $|\rho|$ is close to $1$, the use of an isotropic 9-point stencil is recommended to handle the high correlation scenarios more effectively.\\

Another possibility consists in using a quadratic Hamiltonian approximation as suggested in \cite{bergault2018closed} to reduce the dimensionality of the problem. We can indeed approximate the Hamiltonian functions $H^{0,1}$ and $H^{1,0}$ by quadratic functions:
$$\check{H}^{0,1}: (z,p) \mapsto  \alpha^{0,1}_0(z) + \alpha^{0,1}_1(z) p + \frac 12 \alpha^{0,1}_2(z) p^2 \quad \textrm{and} \quad \check{H}^{1,0}: (z,p) \mapsto  \alpha^{1,0}_0(z) + \alpha^{1,0}_1(z) p + \frac 12 \alpha^{1,0}_2(z) p^2.$$

A natural choice for the functions $\check{H}^{0,1}$ and $\check{H}^{1,0}$ derives from Taylor expansions around $p=0$.\\

We then denote by $\check \theta$ the approximation of $\theta$ associated with the functions $\check{H}^{0,1}$ and $\check{H}^{1,0}$, i.e. $\check \theta$ solves
\begin{equation}
\begin{cases}
 \!&0 = \partial_t \check \theta(t,y, \nu) + \mu y \left( 1 + \partial_y \check \theta(t,y, \nu) \right) - \frac{\gamma}{2} \left(\nu + \bar \lambda {\bar \eta}^2\right) y^2 + \frac 12 \nu y^2 \partial^2_{yy} \check \theta(t,y, \nu)\\
 \!& \qquad + k(\bar \nu - \nu)\partial_\nu \check \theta(t,y, \nu) +  \frac 12 \nu \xi^2 \partial^2_{\nu \nu} \check \theta(t,y, \nu) + \rho \nu y \xi \partial^2_{y\nu} \check \theta(t,y, \nu) \\
 \!& \qquad +\bar \lambda \int_{\eta \in \mathbb R}(\check\theta(t,y(1+\eta), \nu) - \check\theta(t,y, \nu)) \zeta(d\eta)\\ 
\!& \qquad + \text{\scalebox{0.6}[1]{$\bigint$}}_{\!\!\mathbb{R}_{+}^{*}} \left(z\lambda^{0,1}\phi^{0,1}(z) \check H^{0,1} \left(z,\frac{\check \theta(t,y,\nu) -  \check \theta(t,y- z, \nu) }{z}\right) + z\lambda^{1,0}\phi^{1,0}(z) \check H^{1,0} \left(z,\frac{\check \theta(t,y,\nu) -  \check \theta(t,y+z,\nu) }{z}\right) \right)m(dz),\\
\!&\check \theta(T,y,\nu) = 0.
\end{cases}
\label{eqn:quadHJBHB}
\end{equation}

The interest of this quadratic approximation is that Eq. \eqref{eqn:quadHJBHB} has a solution in the form of a quadratic polynomial in $y$, i.e. $\check \theta(t,y, \nu) = - A(t, \nu)y^2 -  B(t, \nu)y - C(t, \nu)$, where $(t,\nu) \mapsto A(t,\nu) \in \mathbb R$, $(t,\nu) \mapsto B(t,\nu) \in \mathbb R$ and $(t,\nu) \mapsto C(t,\nu) \in \mathbb R$ solve a system of PDEs. As the value of $C$ is irrelevant for what follows, we only report here the equations for $A$ and $B$:
\begin{align}\label{PDEsysHB}
\begin{cases}
\partial_tA(t,\nu) &= 2A(t,\nu)^2 \text{\scalebox{0.6}[1]{$\bigint$}}_{\!\!\mathbb{R}_{+}^{*}} z \left(\lambda^{0,1}\phi^{0,1}(z)\alpha^{0,1}_2(z) + \lambda^{1,0}\phi^{1,0}(z)\alpha^{1,0}_2(z) \right)m(dz) - (2\mu + \nu + \bar \lambda {\bar \eta}^2)A(t,\nu)\\
&\qquad - \left( 2\rho \nu \xi +k(\bar \nu - \nu)\right) \partial_\nu A(t,\nu) - \frac 12 \nu \xi^2 \partial^2_{\nu \nu } A(t,\nu)  - \frac{\gamma}{2}\left(\nu + \bar \lambda {\bar \eta}^2\right),\\
\partial_tB(t,\nu)  &= \mu \left(1 - B(t,\nu) \right) -  \left( \rho \nu \xi +k(\bar \nu - \nu)\right) \partial_\nu B(t,\nu) - \frac 12 \nu \xi^2 \partial^2_{\nu \nu } B(t,\nu) \\
&\qquad  + 2A(t,\nu)\text{\scalebox{0.6}[1]{$\bigint$}}_{\!\!\mathbb{R}_{+}^{*}} z \left(\lambda^{0,1}\phi^{0,1}(z)\alpha^{0,1}_1(z) - \lambda^{1,0}\phi^{1,0}(z)\alpha^{1,0}_1(z) \right)m(dz)\\
&\qquad - 2A(t,\nu)^2\text{\scalebox{0.6}[1]{$\bigint$}}_{\!\!\mathbb{R}_{+}^{*}} z^2 \left(\lambda^{0,1}\phi^{0,1}(z)\alpha^{0,1}_2(z) - \lambda^{1,0}\phi^{1,0}(z)\alpha^{1,0}_2(z) \right)m(dz) \\
&\qquad + 2A(t,\nu)B(t,\nu)\text{\scalebox{0.6}[1]{$\bigint$}}_{\!\!\mathbb{R}_{+}^{*}} z \left(\lambda^{0,1}\phi^{0,1}(z)\alpha^{0,1}_2(z) + \lambda^{1,0}\phi^{1,0}(z)\alpha^{1,0}_2(z) \right)m(dz)  ,\\
A(T,\nu) &= 0,\\
B(T,\nu) &= 0.\\
\end{cases}
\end{align}

The PDE system \eqref{PDEsysHB} can be solved numerically far more easily than Eq. \eqref{eqn:HJBHB}. Indeed, the dimensionality of the problem has been reduced and there is no more cross derivatives terms, hence no more geometric problem related to the numerical grid. Furthermore, the system is triangular. One can first solve numerically the equation for $A$ which is a nonlinear parabolic PDE with one state variable whose only nonlinear term is of order $0$. An implicit scheme is therefore perfectly suited to approximate $A$. Once $A$ is approximated numerically, $B$ can be approximated numerically very easily since it is the solution of a linear parabolic PDE with one state variable. Once we have $A$ and $B$, approximations of the optimal markups can be computed by replacing $\theta$ by $\check \theta$ in Eqs. \eqref{optquotes01H} and \eqref{optquotes10H}. We thereby obtain 
\begin{align}\label{quadoptquotes01HB}
 \check \delta^{0,1*}(t,z) = \bar \delta^{0,1} \Big(z, A(t,\nu_t) \left(z-2Y_{t-} \right) -B(t,\nu_t) \Big)
 \end{align}
 and
\begin{align}\label{quadoptquotes10HB}
\check \delta^{1,0*}(t,z) = \bar \delta^{1,0} \Big(z, A(t,\nu_t) \left(z+2Y_{t-} \right) +B(t,\nu_t) \Big).   
 \end{align}

\subsection{A Stein-Stein volatility model with jumps}\label{subSSJ}

\subsubsection{Model, Hamilton-Jacobi-Bellman equation and optimal markups}

Stochastic volatility models were initially introduced in the 1990s as extensions of the Black-Scholes model, designed to capture the volatility smile or skew observed in option prices. The Heston model describes the dynamics of the instantaneous variance using a square-root (CIR) process, whereas the Stein-Stein model utilizes an Ornstein-Uhlenbeck process to model the instantaneous volatility. In the subsequent discussion, we explore an extension of the Stein-Stein model in two directions: (i) the extension \textit{à la} Schöbel and Zhu \cite{schobel1999stochastic}, which permits any correlation coefficient between the Brownian motions driving the price and volatility processes, and (ii) the addition of price jumps.\\

Mathematically, the price process $(S_t)_{t \in \mathbb R_+}$ has the following dynamics:
$$
\begin{cases}
    dS_t = \mu S_t dt + \sigma_t S_t dW^S_t + S_t\int_{\eta \in \mathbb R} \eta \bar J(dt,d\eta),\\
    d\sigma_t = k (\bar \sigma - \sigma_t) dt +  \xi dW^\sigma_t
\end{cases}
$$
where:
\begin{itemize}
\item $\mu \in \mathbb R$ is a known deterministic drift;
\item $(\sigma_t)_{t\in \mathbb R_+}$ is the instantaneous volatility process with values in~$\mathbb R$;
\item $\bar J$ is a marked point process with constant-in-time kernel $\bar \lambda \zeta(d\eta)$ where $\bar \lambda$ is the rate of jumps and $\zeta$ the probability measure of jump sizes in relative value (assumed to be such that $\int_{\eta \in \mathbb R} \eta \zeta(d\eta) = 0$);
\item $k>0$ is the rate of mean reversion of the instantaneous volatility;
\item $\xi>0$ measures the stochasticity of the instantaneous volatility process;
\item $\bar \sigma >0$ is the long-term instantaneous volatility;
\item $\left(W^S_t, W^\sigma_t \right)_{t\in \mathbb R_+}$ is a 2-dimensional standard Brownian motion independent of $\bar J$ with $d\langle W^S, W^\sigma \rangle_t = \rho dt$ for $\rho \in [-1,1]$.\\
\end{itemize}

Like in the Heston-Bates case, it is interesting in this framework  to introduce $Y_t = (q_t^1 - q_0^1) S_t$. We have indeed that $(Y_t, \sigma_t)_{t \in \mathbb R_+}$ is Markovian since
$$ 
dY_t = \mu Y_t dt + \sigma_t Y_t dW^S_t + Y_t \int_{\eta \in \mathbb R} \eta \bar J(dt,d\eta) +  \int_{z \in \mathbb R_+^*}\!\!\!\!\!\!\!\!  z \left(J^{1,0}(dt,dz) -  J^{0,1}(dt,dz) \right)$$ and \eqref{sto_opt_cont} writes
\begin{align}
\sup_{\left(\delta^{0,1},\delta^{1,0}\right) \in \mathcal A}&\mathbb{E}\Bigg[\int\limits_{0}^{T}\Bigg\lbrace \int_{z \in \mathbb R_+^*} \Big(z\delta^{0,1}(t,z)  \Lambda^{0,1}(z,\delta^{0,1}(t,z))+z\delta^{1,0}(t,z)  \Lambda^{1,0}(z,\delta^{1,0}(t,z))  \Big)m(dz)\Bigg\rbrace dt\nonumber\\
& + \int\limits_{0}^{T} \mu Y_t dt  - \frac{\gamma}{2} \int\limits_{0}^{T}\left(\sigma^2_t + \bar \lambda {\bar \eta}^2 \right) Y_t^2 dt  \Bigg], \label{soc_SSJ}
\end{align}
where $\bar \eta = \left(\int_{\eta \in \mathbb R} \eta^2 \zeta(d\eta)\right)^{\frac 12}$.\\

For a given $\gamma$, we associate with \eqref{soc_SSJ} a value function $\theta:[0,T]\times \mathbb R \times \mathbb R \rightarrow \mathbb{R}$ and the following Hamilton-Jacobi-Bellman equation:
\begin{equation}
\begin{cases}
 \!&0 = \partial_t \theta(t,y, \sigma) + \mu y \left( 1 + \partial_y \theta(t,y, \sigma) \right) - \frac{\gamma}{2} \left(\sigma^2 + \bar \lambda {\bar \eta}^2\right) y^2 + \frac 12 \sigma^2 y^2 \partial^2_{yy} \theta(t,y, \sigma)\\
 \!& \qquad + k(\bar \sigma - \sigma)\partial_\sigma \theta(t,y, \sigma) +  \frac 12  \xi^2 \partial^2_{\sigma  \sigma} \theta(t,y,  \sigma) + \rho  \sigma y \xi \partial^2_{y \sigma} \theta(t,y,  \sigma) \\
 \!& \qquad + \bar \lambda \int_{\eta \in \mathbb R}(\theta(t,y(1+\eta), \sigma) - \theta(t,y, \sigma)) \zeta(d\eta) \\
\!& \qquad + \text{\scalebox{0.6}[1]{$\bigint$}}_{\!\!\mathbb{R}_{+}^{*}} \left(z\lambda^{0,1}\phi^{0,1}(z)H^{0,1} \left(z,\frac{\theta(t,y, \sigma) -  \theta(t,y- z, \sigma) }{z}\right) + z\lambda^{1,0}\phi^{1,0}(z)H^{1,0} \left(z,\frac{\theta(t,y, \sigma) -  \theta(t,y+z, \sigma) }{z}\right) \right)m(dz),\\
\!&\theta(T,y, \sigma) = 0,
\end{cases}
\label{eqn:HJBSSJ}
\end{equation} where $H^{0,1}$ and $H^{1,0}$ are defined as above.\\

Using the same ideas as in \cite{gueant2017optimal} and with the same notations as above, we find that the markups that maximize our modified objective function are 
\begin{align}\label{optquotes01SSJ}
 \delta^{0,1*}(t,z) = \bar \delta^{0,1} \left(z, \frac{\theta(t,Y_{t-}, \sigma_t) -  \theta(t,Y_{t-}-z, \sigma_t) }{z} \right)
 \end{align}
 and
\begin{align}\label{optquotes10SSJ}
\delta^{1,0*}(t,z) = \bar \delta^{1,0} \left(z, \frac{\theta(t,Y_{t-}, \sigma_t) -  \theta(t,Y_{t-}+z, \sigma_t) }{z} \right).   
 \end{align}

\subsubsection{Discussion and numerical methods}

Under Stein-Stein dynamics with jumps for the price, computing the optimal markups primarily involves solving the partial integro-differential equation \eqref{eqn:HJBSSJ} to obtain the value function $\theta$. The nature of this equation closely mirrors that of the Heston-Bates case, and the number of state variables remains the same, with the instantaneous volatility $\sigma$ replacing the instantaneous variance $\nu$. Subsequently, one can substitute the current value of $y$ and an estimate of the current value of the instantaneous volatility $\sigma$ (which is not directly observable) into Eqs. \eqref{optquotes01SSJ} and \eqref{optquotes10SSJ} to determine the optimal markups.\\

As with the Heston-Bates model, $\theta$ cannot be found in closed form and must be approximated numerically. The considerations applicable to the Heston-Bates case are also relevant here. In particular, the quadratic Hamiltonian approximation technique introduced in \cite{bergault2018closed} is applicable: we obtain a simpler PDE when approximating the Hamiltonian functions $H^{0,1}$ and $H^{1,0}$ by the quadratic functions
$$\check{H}^{0,1}: (z,p) \mapsto  \alpha^{0,1}_0(z) + \alpha^{0,1}_1(z) p + \frac 12 \alpha^{0,1}_2(z) p^2 \quad \textrm{and} \quad \check{H}^{1,0}: (z,p) \mapsto  \alpha^{1,0}_0(z) + \alpha^{1,0}_1(z) p + \frac 12 \alpha^{1,0}_2(z) p^2.$$

Let us introduce $\check \theta$ the approximation of $\theta$ associated with the functions $\check{H}^{0,1}$ and $\check{H}^{1,0}$. $\check \theta$ then solves
\begin{equation}
\begin{cases}
\!&0 = \partial_t \check \theta(t,y, \sigma) + \mu y \left( 1 + \partial_y \check \theta(t,y, \sigma) \right) - \frac{\gamma}{2} \left(\sigma^2 + 2\bar \lambda \bar \eta^2\right) y^2 + \frac 12 \sigma^2 y^2 \partial^2_{yy} \check \theta(t,y, \sigma)\\
 \!& \qquad + k(\bar \sigma - \sigma)\partial_\sigma \check \theta(t,y, \sigma) +  \frac 12  \xi^2 \partial^2_{\sigma  \sigma} \check \theta(t,y,  \sigma) + \rho  \sigma y \xi \partial^2_{y \sigma} \check \theta(t,y,  \sigma) \\
 \!& \qquad +\bar \lambda \int_{\eta \in \mathbb R}(\check\theta(t,y(1+\eta), \nu) - \check\theta(t,y, \nu)) \zeta(d\eta)  \\
\!& \qquad + \text{\scalebox{0.6}[1]{$\bigint$}}_{\!\!\mathbb{R}_{+}^{*}} \left(z\lambda^{0,1}\phi^{0,1}(z) \check H^{0,1} \left(z,\frac{\check \theta(t,y, \sigma) -  \check \theta(t,y- z, \sigma) }{z}\right) + z\lambda^{1,0}\phi^{1,0}(z) \check H^{1,0} \left(z,\frac{\check \theta(t,y, \sigma) -  \check \theta(t,y+z, \sigma) }{z}\right) \right)m(dz),\\
\!&\check \theta(T,y, \sigma) = 0,
\end{cases}
\label{eqn:quadHJBSSJ}
\end{equation}

As in the Heston-Bates case, Eq. \eqref{eqn:quadHJBSSJ} has a solution in the form of a polynomial of degree $2$ in $y$: $\check \theta(t,y, \sigma) = - A(t, \sigma)y^2 -  B(t, \sigma)y - C(t, \sigma)$, where $(t,\sigma) \mapsto A(t,\sigma) \in \mathbb R$, $(t,\sigma) \mapsto B(t,\sigma) \in \mathbb R$ and $(t,\sigma) \mapsto C(t,\sigma) \in \mathbb R$ solve a system of PDEs. As the value of $C$ is irrelevant for what follows, we only report here the equations for $A$ and $B$:
\begin{align}\label{PDEsysSSJ}
\begin{cases}
\partial_tA(t,\sigma) &= 2A(t,\sigma)^2 \text{\scalebox{0.6}[1]{$\bigint$}}_{\!\!\mathbb{R}_{+}^{*}} z \left(\lambda^{0,1}\phi^{0,1}(z)\alpha^{0,1}_2(z) + \lambda^{1,0}\phi^{1,0}(z)\alpha^{1,0}_2(z) \right)m(dz) - (2\mu + \sigma^2 + \bar \lambda {\bar \eta}^2)A(t,\sigma)\\
&\qquad - \left( 2\rho \sigma \xi +k(\bar \sigma - \sigma)\right) \partial_\sigma A(t,\sigma) - \frac 12  \xi^2 \partial^2_{\sigma \sigma } A(t,\sigma)  - \frac{\gamma}{2}\left(\sigma^2 + \bar \lambda {\bar \eta}^2\right),\\
\partial_tB(t,\sigma)  &= \mu \left(1 - B(t,\sigma) \right) -  \left( \rho \sigma \xi +k(\bar \sigma - \sigma)\right) \partial_\sigma B(t,\sigma) - \frac 12 \xi^2 \partial^2_{\sigma \sigma } B(t,\sigma) \\
&\qquad  + 2A(t,\sigma)\text{\scalebox{0.6}[1]{$\bigint$}}_{\!\!\mathbb{R}_{+}^{*}} z \left(\lambda^{0,1}\phi^{0,1}(z)\alpha^{0,1}_1(z) - \lambda^{1,0}\phi^{1,0}(z)\alpha^{1,0}_1(z) \right)m(dz) \\
&\qquad - 2A(t,\sigma)^2\text{\scalebox{0.6}[1]{$\bigint$}}_{\!\!\mathbb{R}_{+}^{*}} z^2 \left(\lambda^{0,1}\phi^{0,1}(z)\alpha^{0,1}_2(z) - \lambda^{1,0}\phi^{1,0}(z)\alpha^{1,0}_2(z) \right)m(dz)\\
&\qquad + 2A(t,\sigma)B(t,\sigma)\text{\scalebox{0.6}[1]{$\bigint$}}_{\!\!\mathbb{R}_{+}^{*}} z \left(\lambda^{0,1}\phi^{0,1}(z)\alpha^{0,1}_2(z) + \lambda^{1,0}\phi^{1,0}(z)\alpha^{1,0}_2(z) \right)m(dz)  ,\\
A(T,\sigma) &= 0,\\
B(T,\sigma) &= 0.\\
\end{cases}
\end{align}
The PDE system \eqref{PDEsysSSJ} can be approximated very easily using an implicit scheme. Once $A$ and $B$ are obtained, approximations of the optimal strategies can be computed by replacing $\theta$ by $\check \theta$ in Eqs. \eqref{optquotes01SSJ} and \eqref{optquotes10SSJ}. We thereby obtain 
\begin{align}\label{quadoptquotes01SSJ}
 \check \delta^{0,1*}(t,z) = \bar \delta^{0,1} \Big(z, A(t,\sigma_t) \left(z-2Y_{t-} \right) -B(t,\sigma_t) \Big)
 \end{align}
 and
\begin{align}\label{quadoptquotes10SSJ}
\check \delta^{1,0*}(t,z) = \bar \delta^{1,0} \Big(z, A(t,\sigma_t) \left(z+2Y_{t-} \right) +B(t,\sigma_t) \Big).   
 \end{align}

\subsection{Beyond stochastic volatility models: the use of a simple Markovian Hawkes framework}\label{subH}

\subsubsection{Model, Hamilton-Jacobi-Bellman equation and optimal markups}

Beyond stochastic volatility models, there exists a rich landscape of alternative frameworks for price dynamics, especially those emerging from the field of market microstructure. One prominent category of these models extensively utilizes Hawkes processes to capture the self-exciting nature of market events and their influence on asset prices. These models provide a distinct perspective on price evolution in financial markets and can more effectively capture certain stylized facts, like time reversal asymmetries, compared to traditional stochastic volatility models (see \cite{bouchaud2018trades}).\\

In what follows, we propose one of the simplest pure jump models, which relies on Hawkes processes for the price dynamics. Mathematically, the dynamics of $(S_t)_{t \in \mathbb{R}_+}$ writes as follows:\footnote{In this model, we consider zero drift.}
$$dS_t =  S_t\int_{\eta \in \mathbb R} \eta \bar J(dt,d\eta)$$
where $\bar J$ is a marked point process with dynamic kernel $\bar \lambda_t \zeta(d\eta)$ with $\zeta$ the probability measure of jump sizes in relative value (assumed to be such that $\int_{\eta \in \mathbb R} \eta \zeta(d\eta) = 0$) and $(\bar \lambda_t)_{t \in \mathbb R_+}$ following the dynamics\footnote{Power law kernels may be more realistic than exponential ones, but they do not lead to Markovian setups.}
$$d\bar\lambda_t = -\beta(\bar\lambda_t - \bar\lambda_\infty) dt + \beta \frak{n} \int_{\eta \in \mathbb R} \bar J(dt,d\eta)$$ with $\bar \lambda_0$ higher than the long-term limit $\bar\lambda_\infty$ in the absence of jumps, $\beta > 0$ controlling the memory of the process and $\frak{n} \in (0,1)$ corresponding to the percentage of price moves that are purely endogenous.\\

As above, we introduce \(Y_t = (q_t^1 - q_0^1) S_t\) and observe that the pair \((Y_t, \bar{\lambda}_t)_{t \in \mathbb{R}_+}\) is Markovian since
$$ 
dY_t = Y_t \int_{\eta \in \mathbb R} \eta \bar J(dt,d\eta) +  \int_{z \in \mathbb R_+^*}\!\!\!\!\!\!\!\!  z \left(J^{1,0}(dt,dz) -  J^{0,1}(dt,dz) \right).$$ and \eqref{sto_opt_cont} writes therefore as
\begin{align}
\sup_{\left(\delta^{0,1},\delta^{1,0}\right) \in \mathcal A}&\mathbb{E}\Bigg[\int\limits_{0}^{T}\Bigg\lbrace \int_{z \in \mathbb R_+^*} \Big(z\delta^{0,1}(t,z)  \Lambda^{0,1}(z,\delta^{0,1}(t,z))+z\delta^{1,0}(t,z)  \Lambda^{1,0}(z,\delta^{1,0}(t,z))  \Big)m(dz)\Bigg\rbrace dt\nonumber\\
&  - \frac{\gamma}{2} \int\limits_{0}^{T}\bar\lambda_t {\bar \eta}^2 Y_t^2 dt  \Bigg], \label{soc_H}
\end{align}
where $\bar \eta = \left(\int_{\eta \in \mathbb R} \eta^2 \zeta(d\eta)\right)^{\frac 12}$.\\

For a given $\gamma$, we associate with \eqref{soc_H} a value function $\theta:[0,T]\times \mathbb R \times \mathbb R_+^* \rightarrow \mathbb{R}$ and the following Hamilton-Jacobi-Bellman equation:
\begin{equation}
\begin{cases}
 \!&0 = \partial_t \theta(t,y, \bar\lambda) - \frac{\gamma}{2} \bar\lambda {\bar \eta}^2 y^2 
 -\beta(\bar\lambda - \bar\lambda_\infty)\partial_{\bar\lambda} \theta(t,y, \bar\lambda)+ \bar\lambda \int_{\eta \in \mathbb R}(\theta(t,y(1+\eta), \bar\lambda + \beta \frak{n}) - \theta(t,y, \bar\lambda)) \zeta(d\eta) \\
\!& \quad + \text{\scalebox{0.6}[1]{$\bigint$}}_{\!\!\mathbb{R}_{+}^{*}} \left(z\lambda^{0,1}\phi^{0,1}(z)H^{0,1} \left(z,\frac{\theta(t,y, \bar\lambda) -  \theta(t,y- z, \bar\lambda) }{z}\right) + z\lambda^{1,0}\phi^{1,0}(z)H^{1,0}H^{1,0} \left(z,\frac{\theta(t,y, \bar\lambda) -  \theta(t,y+z, \bar\lambda) }{z}\right) \right)m(dz),\\
\!&\theta(T,y, \lambda) = 0,
\end{cases}
\label{eqn:HJBH}
\end{equation} where $H^{0,1}$ and $H^{1,0}$ are defined as above.\\

Using the same ideas as in \cite{gueant2017optimal} and with the same notations as above, we find that the markups that maximize our modified objective function are 
\begin{align}\label{optquotes01Hawkes}
 \delta^{0,1*}(t,z) = \bar \delta^{0,1} \left(z, \frac{\theta(t,Y_{t-}, \bar\lambda_{t-}) -  \theta(t,Y_{t-}-z, \bar\lambda_{t-}) }{z} \right)
 \end{align}
 and
\begin{align}\label{optquotes10Hawkes}
\delta^{1,0*}(t,z) = \bar \delta^{1,0} \left(z, \frac{\theta(t,Y_{t-}, \bar\lambda_{t-}) -  \theta(t,Y_{t-}+z, \bar\lambda_{t-}) }{z} \right).   
 \end{align}

\subsubsection{Discussion and numerical methods}

Under the Hawkes dynamics specified for the price, computing the optimal markups primarily requires solving a partial integro-differential equation  -- here \eqref{eqn:HJBH} -- to derive the value function $\theta$. This function depends on time and two state variables: the spread to Hodl $y$ and the intensity $\bar\lambda$. Subsequently, one can substitute the current values of $y$ and $\bar\lambda$ into Eqs. \eqref{optquotes01Hawkes} and \eqref{optquotes10Hawkes} to determine the optimal markups. It is important to note that there are two different approaches for determining the current value of $\bar\lambda$. The first approach starts with an initial value $\bar \lambda_0$ (whose influence diminishes over time) and treats $\bar\lambda$ as an observable variable that evolves in real-time with market dynamics. The second approach uses statistical methods to estimate the current value of $\bar\lambda$ based on recent historical price data. In this scenario, $\bar\lambda$ is not treated as an observable variable, but rather similarly to how instantaneous volatility was treated in the case of stochastic volatility models.\\

It is possible to approximate $\theta$ numerically and subsequently make a market making strategy rely on the resolution of Eq. \eqref{eqn:HJBH}. However, to reduce the dimensionality of equations by $1$, one can again use the quadratic Hamiltonian approximation technique, although it leads to more complex equations than in the case of stochastic volatility models because of the nonlocal term in $\bar \lambda$. Let us indeed define $\check \theta$ as the approximation of $\theta$ associated with the functions $$\check{H}^{0,1}: (z,p) \mapsto  \alpha^{0,1}_0(z) + \alpha^{0,1}_1(z) p + \frac 12 \alpha^{0,1}_2(z) p^2 \quad \textrm{and} \quad \check{H}^{1,0}: (z,p) \mapsto  \alpha^{1,0}_0(z) + \alpha^{1,0}_1(z) p + \frac 12 \alpha^{1,0}_2(z) p^2.$$
$\check \theta$ solves
\begin{equation}
\begin{cases}
\!&0 = \partial_t \check\theta(t,y, {\bar \lambda}) - \frac{\gamma}{2} {\bar \lambda} {\bar \eta}^2 y^2 
 -\beta({\bar \lambda} - {\bar \lambda}_\infty)\partial_{\bar \lambda} \check\theta(t,y, {\bar \lambda})+ {\bar \lambda} \int_{\eta \in \mathbb R}(\check\theta(t,y(1+\eta), {\bar \lambda} + \beta \frak{n}) - \check\theta(t,y, {\bar \lambda})) \zeta(d\eta) \\
\!& \qquad + \text{\scalebox{0.6}[1]{$\bigint$}}_{\!\!\mathbb{R}_{+}^{*}} \left(z\lambda^{0,1}\phi^{0,1}(z) \check H^{0,1} \left(z,\frac{\check \theta(t,y, {\bar \lambda}) -  \check \theta(t,y- z, {\bar \lambda}) }{z}\right) + z\lambda^{1,0}\phi^{1,0}(z) \check H^{1,0} \left(z,\frac{\check \theta(t,y, {\bar \lambda}) -  \check \theta(t,y+z, {\bar \lambda}) }{z}\right) \right)m(dz),\\
\!&\check \theta(T,y, {\bar \lambda}) = 0.
\end{cases}
\label{eqn:quadHJBH}
\end{equation}

As above, Eq. \eqref{eqn:quadHJBH} has a solution of the form $\check \theta(t,y, {\bar \lambda}) = - A(t, {\bar \lambda})y^2 -  B(t, {\bar \lambda})y - C(t, {\bar \lambda})$, where $(t,{\bar \lambda}) \mapsto A(t,{\bar \lambda}) \in \mathbb R$, $(t,{\bar \lambda}) \mapsto B(t,{\bar \lambda}) \in \mathbb R$ and $(t,{\bar \lambda}) \mapsto C(t,{\bar \lambda}) \in \mathbb R$ solve a system of PDEs, but this time with nonlocal terms. As the value of $C$ is irrelevant for what follows, we only report the equations for $A$ and $B$:
\begin{align}\label{PDEsysH}
\begin{cases}
\partial_tA(t,{\bar \lambda}) &= 2A(t,{\bar \lambda})^2 \text{\scalebox{0.6}[1]{$\bigint$}}_{\!\!\mathbb{R}_{+}^{*}} z \left(\lambda^{0,1} \phi^{0,1}(z) \alpha^{0,1}_2(z) + \lambda^{1,0} \phi^{1,0}(z) \alpha^{1,0}_2(z) \right)m(dz) - \frac{\gamma}{2}{\bar \lambda} {\bar \eta}^2\\
&\quad + \beta({\bar \lambda} - {\bar \lambda}_\infty)\partial_{\bar \lambda} A(t,{\bar \lambda}) - {\bar \lambda}\left(A(t,{\bar \lambda} + \beta \frak{n}) (1 +{\bar\eta}^2) - A(t,{\bar \lambda})\right),\\ 
\partial_tB(t,{\bar \lambda})  &= \beta({\bar \lambda} - {\bar \lambda}_\infty)\partial_{\bar \lambda} B(t,{\bar \lambda}) - {\bar \lambda}\left(B(t,{\bar \lambda} + \beta \frak{n}) - B(t,{\bar \lambda})\right)\\
&\qquad  + 2A(t,{\bar \lambda})\text{\scalebox{0.6}[1]{$\bigint$}}_{\!\!\mathbb{R}_{+}^{*}} z \left(\lambda^{0,1} \phi^{0,1}(z) \alpha^{0,1}_1(z) - \lambda^{1,0} \phi^{1,0}(z) \alpha^{1,0}_1(z) \right)m(dz)\\
&\qquad - 2A(t,{\bar \lambda})^2\text{\scalebox{0.6}[1]{$\bigint$}}_{\!\!\mathbb{R}_{+}^{*}} z^2 \left(\lambda^{0,1} \phi^{0,1}(z) \alpha^{0,1}_2(z) - \lambda^{1,0} \phi^{1,0}(z) \alpha^{1,0}_2(z) \right)m(dz) \\
&\qquad + 2A(t,{\bar \lambda})B(t,{\bar \lambda})\text{\scalebox{0.6}[1]{$\bigint$}}_{\!\!\mathbb{R}_{+}^{*}} z \left(\lambda^{0,1} \phi^{0,1}(z) \alpha^{0,1}_2(z) + \lambda^{1,0} \phi^{1,0}(z) \alpha^{1,0}_2(z) \right)m(dz)  ,\\
A(T,{\bar \lambda}) &= 0,\\
B(T,{\bar \lambda}) &= 0.\\
\end{cases}
\end{align}
The triangular system of PDEs with nonlocal terms \eqref{PDEsysH} can be approximated using an implicit scheme. Once $A$ and $B$ are obtained, approximations of the optimal strategies can be computed by replacing $\theta$ by $\check \theta$ in Eqs.~\eqref{optquotes01Hawkes} and \eqref{optquotes10Hawkes}. We thereby obtain 
\begin{align}\label{quadoptquotes01H}
 \check \delta^{0,1*}(t,z) = \bar \delta^{0,1} \Big(z, A(t,{\bar \lambda}_{t-}) \left(z-2Y_{t-} \right) -B(t,{\bar \lambda}_{t-}) \Big)
 \end{align}
 and
\begin{align}\label{quadoptquotes10H}
\check \delta^{1,0*}(t,z) = \bar \delta^{1,0} \Big(z, A(t,{\bar \lambda}_{t-}) \left(z+2Y_{t-} \right) +B(t,{\bar \lambda}_{t-}) \Big).   
 \end{align}
 
\subsection{Extension to quadratic Hawkes processes}
\label{subZH}

\subsubsection{Model, Hamilton-Jacobi-Bellman equation and optimal markups}

Extensions of Hawkes processes have been proposed in the market microstructure literature to model intraday effects such as the Zumbach effect. In particular, \cite{ZH} introduced what they termed Z-Hawkes, which are low-rank quadratic Hawkes processes. Inspired by this approach, we consider in this section that the price process $(S_t)_{t \in \mathbb{R}_+}$ is a pure jump process,\footnote{Here again, we consider zero drift.} i.e.
$$dS_t =  S_t\int_{\eta \in \mathbb R} \eta \bar J(dt,d\eta)$$
where $\bar J$ is a marked point process with dynamic kernel $\bar \lambda_t \zeta(d\eta)$ where $\zeta$ is the probability measure of jump sizes in relative value (assumed to be such that $\int_{\eta \in \mathbb R} \eta \zeta(d\eta) = 0$), but where the process $(\bar \lambda_t)_{t \in \mathbb R_+}$ is now defined by
$$\bar \lambda_t = \bar \lambda_\infty + h_t + \xi_t^2$$
where
\begin{eqnarray*}
dh_t &=& -\kappa h_t dt + \kappa \frak{n}_H \int_{\eta \in \mathbb R} \bar J(dt,d\eta),\\
d\xi_t &=& - \omega \xi_t dt + \sqrt{2\frak{n}_Z \omega} \int_{\eta \in \mathbb R} \text{sign}(\eta) \bar J(dt,d\eta)
\end{eqnarray*}
with $\kappa, \omega, \frak{n}_H, \frak{n}_Z  >0$ and $\frak{n}_H + \frak{n}_Z < 1$.\footnote{This model is an extension of the model of Section \ref{subH} since we recover it when $\xi_0 = \omega = 0$.}\\

Introducing as above $Y_t = (q_t^1 - q_0^1) S_t$, we see that
$$ 
dY_t = Y_t \int_{\eta \in \mathbb R} \eta \bar J(dt,d\eta) +  \int_{z \in \mathbb R_+^*}\!\!\!\!\!\!\!\!  z \left(J^{1,0}(dt,dz) -  J^{0,1}(dt,dz) \right).$$ Subsequently, the triplet $(Y_t, h_t, \xi_t)_{t \in \mathbb R_+}$ is Markovian.\\

The stochastic optimal control problems \eqref{sto_opt_cont} then write
\begin{align}
\sup_{\left(\delta^{0,1},\delta^{1,0}\right) \in \mathcal A}&\mathbb{E}\Bigg[\int\limits_{0}^{T}\Bigg\lbrace \int_{z \in \mathbb R_+^*} \Big(z\delta^{0,1}(t,z)  \Lambda^{0,1}(z,\delta^{0,1}(t,z))+z\delta^{1,0}(t,z)  \Lambda^{1,0}(z,\delta^{1,0}(t,z))  \Big)m(dz)\Bigg\rbrace dt\nonumber\\
&  - \frac{\gamma}{2} \int\limits_{0}^{T}(\bar \lambda_\infty + h_t + \xi_t^2) {\bar \eta}^2 Y_t^2 dt  \Bigg], \label{soc_ZH}
\end{align}
where $\bar \eta = \left(\int_{\eta \in \mathbb R} \eta^2 \zeta(d\eta)\right)^{\frac 12}$.\\

For a given $\gamma$, we associate with \eqref{soc_ZH} a value function $\theta:[0,T]\times \mathbb R \times \mathbb R_+^* \times \mathbb R  \rightarrow \mathbb{R}$ and the following Hamilton-Jacobi-Bellman equation:
\begin{equation}
\begin{cases}
 \!&0 = \partial_t \theta(t,y, h, \xi) - \frac{\gamma}{2} (\lambda_\infty + h + \xi^2) {\bar \eta}^2 y^2 
 -\kappa h\partial_h \theta(t,y,h,\xi) -\omega \xi\partial_\xi \theta(t,y,h,\xi)\\
 &\qquad + (\bar \lambda_\infty + h + \xi^2) \left(\int_{\eta >0}(\theta(t,y(1+\eta), h + \kappa \frak{n}_H, \xi +  \sqrt{2\frak{n}_Z \omega}) - \theta(t,y, \lambda)) \zeta(d\eta)\right.\\
 &\qquad \qquad\qquad\qquad\quad  \left. + \int_{\eta <0}(\theta(t,y(1+\eta), h + \kappa \frak{n}_H, \xi -  \sqrt{2\frak{n}_Z \omega}) - \theta(t,y, \lambda)) \zeta(d\eta)\right)  \\
\!& \qquad + \text{\scalebox{0.6}[1]{$\bigint$}}_{\!\!\mathbb{R}_{+}^{*}} \left(z\lambda^{0,1} \phi^{0,1}(z)H^{0,1} \left(z,\frac{\theta(t,y, \lambda) -  \theta(t,y- z, \lambda) }{z}\right) + z\lambda^{1,0} \phi^{1,0}(z)H^{1,0} \left(z,\frac{\theta(t,y, \lambda) -  \theta(t,y+z, \lambda) }{z}\right) \right)m(dz),\\
\!&\theta(T,y, \lambda) = 0,
\end{cases}
\label{eqn:HJBZH}
\end{equation} where $H^{0,1}$ and $H^{1,0}$ are defined as above.\\

Using the same ideas as in \cite{gueant2017optimal} and with the same notations as above, we find that the optimal markups are given by 
\begin{align}\label{optquotes01ZHawkes}
 \delta^{0,1*}(t,z) = \bar \delta^{0,1} \left(z, \frac{\theta(t,Y_{t-}, h_{t-}, \xi_{t-}) -  \theta(t,Y_{t-}-z, h_{t-}, \xi_{t-}) }{z} \right)
 \end{align}
 and
\begin{align}\label{optquotes10ZHawkes}
\delta^{1,0*}(t,z) = \bar \delta^{1,0} \left(z, \frac{\theta(t,Y_{t-}, h_{t-}, \xi_{t-}) -  \theta(t,Y_{t-}+z, h_{t-}, \xi_{t-}) }{z} \right).   
 \end{align}

\subsubsection{Discussion and numerical methods}

Under the specified dynamics for the price, computing the optimal markups largely entails solving the partial integro-differential equation \eqref{eqn:HJBZH} to determine the value function $\theta$, which is dependent on time and three state variables: the spread to Hodl $y$, and the intensity drivers $h$ and $\xi$.\footnote{Similar to the intensity in the classical Hawkes model, $h$ and $\xi$ can be regarded as observable variables.} Due to its high dimensionality, this equation poses significant challenges for numerical solutions. However, the dimensionality can be reduced by one using the quadratic Hamiltonian approximation technique, though this approach still results in more complex equations compared to those found in stochastic volatility models. Let us indeed define $\check \theta$ as the approximation of $\theta$ associated with the functions $$\check{H}^{0,1}: (z,p) \mapsto  \alpha^{0,1}_0(z) + \alpha^{0,1}_1(z) p + \frac 12 \alpha^{0,1}_2(z) p^2 \quad \textrm{and} \quad \check{H}^{1,0}: (z,p) \mapsto  \alpha^{1,0}_0(z) + \alpha^{1,0}_1(z) p + \frac 12 \alpha^{1,0}_2(z) p^2.$$
$\check \theta$ then solves
\begin{equation}
\begin{cases}
\!&0 = \partial_t \check\theta(t,y, h, \xi) - \frac{\gamma}{2} (\lambda_\infty + h + \xi^2) {\bar \eta}^2 y^2 
 -\kappa h\partial_h \check\theta(t,y,h,\xi) -\omega \xi\partial_\xi \check\theta(t,y,h,\xi)\\
 &\quad + (\bar \lambda_\infty + h + \xi^2) \left(\int_{\eta >0}(\check\theta(t,y(1+\eta), h + \kappa \frak{n}_H, \xi +  \sqrt{2\frak{n}_Z \omega}) - \check\theta(t,y, \lambda)) \zeta(d\eta)\right.\\
 &\quad \qquad\qquad\qquad  \left. + \int_{\eta <0}(\check\theta(t,y(1+\eta), h + \kappa \frak{n}_H, \xi -  \sqrt{2\frak{n}_Z \omega}) - \check\theta(t,y, \lambda)) \zeta(d\eta)\right)  \\
\!& \quad + \text{\scalebox{0.6}[1]{$\bigint$}}_{\!\!\mathbb{R}_{+}^{*}} \left(z\lambda^{0,1}\phi^{0,1}(z) \check H^{0,1} \left(z,\frac{\check \theta(t,y, h, \xi) -  \check \theta(t,y- z, h, \xi) }{z}\right) + z\lambda^{1,0}\phi^{1,0}(z) \check H^{1,0} \left(z,\frac{\check \theta(t,y, h, \xi) -  \check \theta(t,y+z, h, \xi) }{z}\right) \right)m(dz),\\
\!&\check \theta(T,y, h, \xi) = 0.
\end{cases}
\label{eqn:quadHJBZH}
\end{equation}

As above, Eq. \eqref{eqn:quadHJBZH} has a quadratic solution of the form $\check \theta(t,y, h, \xi) = - A(t, h, \xi)y^2 -  B(t, h, \xi)y - C(t, h, \xi)$, where $(t,h, \xi) \mapsto A(t,h, \xi) \in \mathbb R$, $(t,h, \xi) \mapsto B(t,h, \xi) \in \mathbb R$ and $(t,h, \xi) \mapsto C(t,h, \xi) \in \mathbb R$ solve a system of PDEs with nonlocal terms. As the value of $C$ is irrelevant for what follows, we only report the equations for $A$ and~$B$:

\begin{align}\label{PDEsysZH}
\begin{cases}
\partial_tA(t,h, \xi) &= 2A(t,h, \xi)^2 \text{\scalebox{0.6}[1]{$\bigint$}}_{\!\!\mathbb{R}_{+}^{*}} z \left(\lambda^{0,1} \phi^{0,1}(z)\alpha^{0,1}_2(z) + \lambda^{1,0} \phi^{1,0}(z)\alpha^{1,0}_2(z) \right)m(dz)  - \frac{\gamma}{2}(\lambda_\infty + h + \xi^2) {\bar \eta}^2\\
&\quad + \kappa h\partial_h A(t,h, \xi) + \omega \xi\partial_\xi A(t,h, \xi)\\
&\quad - (\lambda_\infty + h + \xi^2)\left( \left(\int_{\eta > 0}(1 +\eta)^2 \zeta(d\eta)\right) A(t,h + \kappa \frak{n}_H, \xi +  \sqrt{2\frak{n}_Z \omega})\right.\\
&\qquad  \qquad  \qquad\qquad  + \left.\left(\int_{\eta < 0}(1 +\eta)^2 \zeta(d\eta)\right)A(t,h + \kappa \frak{n}_H, \xi -  \sqrt{2\frak{n}_Z \omega}) - A(t,h, \xi)\right),\\
\partial_tB(t,h, \xi)  &= \kappa h\partial_h B(t,h, \xi) + \omega \xi\partial_\xi B(t,h, \xi)\\ 
&\quad - (\lambda_\infty + h + \xi^2)\left( \left(\int_{\eta > 0}(1 +\eta) \zeta(d\eta)\right) B(t,h + \kappa \frak{n}_H, \xi +  \sqrt{2\frak{n}_Z \omega})\right.\\
&\qquad  \qquad  \qquad\qquad \left. + \left(\int_{\eta < 0}(1 +\eta) \zeta(d\eta)\right)B(t,h + \kappa \frak{n}_H, \xi -  \sqrt{2\frak{n}_Z \omega}) - B(t,h, \xi)\right)\\
&\qquad  + 2A(t,h, \xi)\text{\scalebox{0.6}[1]{$\bigint$}}_{\!\!\mathbb{R}_{+}^{*}} z \left(\lambda^{0,1} \phi^{0,1}(z)\alpha^{0,1}_1(z) - \lambda^{1,0} \phi^{1,0}(z)\alpha^{1,0}_1(z) \right)m(dz)\\
&\qquad - 2A(t,h, \xi)^2\text{\scalebox{0.6}[1]{$\bigint$}}_{\!\!\mathbb{R}_{+}^{*}} z^2 \left(\lambda^{0,1} \phi^{0,1}(z)\alpha^{0,1}_2(z) - \lambda^{1,0} \phi^{1,0}(z)\alpha^{1,0}_2(z) \right)m(dz) \\
&\qquad + 2A(t,h, \xi)B(t,h, \xi)\text{\scalebox{0.6}[1]{$\bigint$}}_{\!\!\mathbb{R}_{+}^{*}} z \left(\lambda^{0,1} \phi^{0,1}(z)\alpha^{0,1}_2(z) + \lambda^{1,0} \phi^{1,0}(z)\alpha^{1,0}_2(z) \right)m(dz),\\
A(T,h, \xi) &= B(T,h, \xi) = 0.
\end{cases}
\end{align}

The triangular system of PDEs with nonlocal terms \eqref{PDEsysZH} can be approximated using an implicit scheme. Once $A$ and $B$ are obtained, approximations of the optimal strategies can be computed by replacing $\theta$ by $\check \theta$ in Eqs. \eqref{optquotes01ZHawkes} and \eqref{optquotes10ZHawkes}. We thereby obtain 
\begin{align}\label{quadoptquotes01ZH}
 \check \delta^{0,1*}(t,z) = \bar \delta^{0,1} \Big(z, A(t, h_{t-}, \xi_{t-}) \left(z-2Y_{t-} \right) -B(t,h_{t-}, \xi_{t-}) \Big)
 \end{align}
 and
\begin{align}\label{quadoptquotes10ZH}
\check \delta^{1,0*}(t,z) = \bar \delta^{1,0} \Big(z, A(t,h_{t-}, \xi_{t-}) \left(z+2Y_{t-} \right) +B(t,h_{t-}, \xi_{t-}) \Big).   
 \end{align}

\subsection{Conclusion on advanced price dynamics models}

In this section, we have addressed two stochastic volatility models from the derivatives pricing literature and two Hawkes-based models from the market microstructure literature. Utilizing these models to compute optimal quotes for a price-aware automated market maker (AMM) requires solving a partial differential equation (PDE). Once the quadratic Hamiltonian approximation technique is applied, for the first three models, these PDEs involve time and one spatial dimension, enabling the use of fast numerical methods. However, this task becomes more complex with models such as the quadratic Hawkes model, which presents greater computational challenges to be used in practice.\\

Like all automated market makers, price-aware AMMs can offer swaps across a vast spectrum of cryptocurrencies. The selection of an appropriate model for determining price dynamics is crucial and depends on the specific cryptocurrency pair involved. For instance, pools involving a stablecoin and a major cryptocurrency might utilize a different model than pools for two independent cryptocurrencies, or pairs of cryptocurrencies that are highly correlated due to shared risk factors.\\

For each cryptocurrency pair, there is a significant trade-off between the simplicity of the model and its capacity to realistically capture market dynamics. This trade-off not only influences the computational efficiency of the AMM but also affects its effectiveness in accurately reflecting the inherent market dynamics of each asset pair, and thus its ability to manage risk appropriately.

\section{Efficient market making strategies with varying liquidity}
\label{liq}

\subsection{Introduction}

In most academic papers on market making, liquidity at the bid and the offer is assumed to be constant. However, this simplification can be detrimental when fluctuations in liquidity are significant. In reality, liquidity experiences fluctuations: there are periods of high liquidity as well as periods of low liquidity. Moreover, there may be times when buyers significantly outnumber sellers, or vice versa, underscoring the fundamental role of market makers. Despite the practical relevance of these variations, they have largely been overlooked in the existing literature on market making. Notable exceptions include the MRR model discussed in \cite{bouchaud2018trades}, which explores autocorrelation in the signs of trades, and the more recent paper \cite{bergault2023modeling}, which employs a Markov-modulated Poisson process to capture these dynamics.\\

Our objective in this section is twofold. First, we examine the use of Hawkes processes for modeling liquidity dynamics -- an approach that might seem natural to many scholars and practitioners. However, this method leads to equations of high dimensionality, thereby making the use of Hawkes processes for modeling liquidity impractical for the stochastic optimal control problems at hand. Second, we suggest that Markov-modulated Poisson processes represent a compelling alternative, capable of capturing the stochastic dynamics of liquidity while avoiding the dimensionality issues that hamper models based on Hawkes processes.\footnote{Deterministic fluctuations of liquidity, such as those linked to intraday patterns, can be incorporated without increasing modeling complexity.}

\subsection{Hawkes models for liquidity and the curse of dimensionality}

\subsubsection{Model, Hamilton-Jacobi-Bellman equation and optimal markups}

Throughout Section \ref{sec_prices}, we assumed $\lambda_t^{0,1}$ and $\lambda_t^{1,0}$ were constant. Instead, it is natural to assume that $(\lambda_t^{0,1})_{t  \in \mathbb R+}$ and $(\lambda_t^{1,0})_{t  \in \mathbb R+}$ are stochastic processes impacted by trades themselves. In a typical Hawkes setting,\footnote{More complex settings can be considered but they all suffer from the curse of dimensionality.} we can assume that
$$
\begin{cases}
d\lambda_t^{0,1} = -\kappa (\lambda_t^{0,1} - \lambda_\infty^{0,1}) dt + \kappa \frak{m}\int_{z \in \mathbb R} J^{0,1}(dt,dz)\\
d\lambda_t^{1,0} = -\kappa (\lambda_t^{1,0} - \lambda_\infty^{1,0}) dt + \kappa \frak{m}\int_{z \in \mathbb R} J^{1,0}(dt,dz)\\
\end{cases}$$
with $\kappa >0$ and $0<\frak{m}<1$.\\

Assuming that the price process is a geometric Brownian motion with drift,\footnote{The extensions of Section \ref{sec_prices} can of course be considered but we go back to the simplest price model for the sake of parsimony.} i.e. $dS_t = \mu S_t  dt + \sigma S_t dW_t$, $Y_t = (q_t^1 - q_0^1) S_t$ has the  dynamics 
$$dY_t = \mu Y_t dt + \sigma Y_t dW_t +  \int_{z \in \mathbb R_+^*}\!\!\!\!\!\!\!\!  z \left(J^{1,0}(dt,dz) -  J^{0,1}(dt,dz) \right)$$ and we see therefore that $(Y_t, \lambda_t^{0,1}, \lambda_t^{1,0})_{t \in \mathbb R_+}$ is Markovian.\\

The stochastic optimal control problems \eqref{sto_opt_cont} write therefore
\begin{align}
\sup_{\left(\delta^{0,1},\delta^{1,0}\right) \in \mathcal A}&\mathbb{E}\Bigg[\int\limits_{0}^{T}\Bigg\lbrace \int_{z \in \mathbb R_+^*} \Big(z\delta^{0,1}(t,z)  \Lambda^{0,1}(t,z,\delta^{0,1}(t,z))+z\delta^{1,0}(t,z)  \Lambda^{1,0}(t,z,\delta^{1,0}(t,z))  \Big)m(dz)\Bigg\rbrace dt\nonumber\\
& + \int\limits_{0}^{T} \mu Y_t dt  - \frac{\gamma}{2} \sigma^2 \int\limits_{0}^{T} Y_t^2 dt  \Bigg]. \label{soc_liqH}
\end{align}

For a given $\gamma$, we associate with \eqref{soc_liqH} a value function $\theta:[0,T]\times \mathbb R \times \mathbb R_+^* \times \mathbb R_+^* \rightarrow \mathbb{R}$ and the following Hamilton-Jacobi-Bellman equation:
\begin{equation}
\begin{cases}
 \!&0 = \partial_t \theta(t,y, \lambda^{0,1}, \lambda^{1,0}) + \mu y \left( 1 + \partial_y \theta(t,y, \lambda^{0,1}, \lambda^{1,0}) \right) - \frac{\gamma}{2} \sigma^2 y^2 + \frac 12 \sigma^2 y^2 \partial^2_{yy} \theta(t,y, \lambda^{0,1}, \lambda^{1,0})\\
\!& \quad -\kappa (\lambda^{0,1} - \lambda_\infty^{0,1}) \partial_{\lambda^{0,1}} \theta(t,y, \lambda^{0,1}, \lambda^{1,0}) -\kappa (\lambda^{1,0} - \lambda_\infty^{1,0}) \partial_{\lambda^{1,0}} \theta(t,y, \lambda^{0,1}, \lambda^{1,0})\\
\!& \qquad + \text{\scalebox{0.6}[1]{$\bigint$}}_{\!\!\mathbb{R}_{+}^{*}} \left(z\lambda^{0,1} \phi^{0,1}(z)H^{0,1} \left(z,\frac{\theta(t,y, \lambda^{0,1}, \lambda^{1,0}) -  \theta(t,y- z, \lambda^{0,1} + \kappa \frak{m}, \lambda^{1,0}) }{z}\right)\right.\\
\!&\qquad\qquad\quad \left. + z\lambda^{1,0} \phi^{1,0}(z)H^{1,0} \left(z,\frac{\theta(t,y, \lambda^{0,1}, \lambda^{1,0}) -  \theta(t,y+z, \lambda^{0,1}, \lambda^{1,0}+\kappa \frak{m}) }{z}\right) \right)m(dz),\\
\!&\theta(T,y, \lambda^{0,1}, \lambda^{1,0}) = 0,
\end{cases}
\label{eqn:HJBliqH}
\end{equation} where $H^{0,1}$ and $H^{1,0}$ are defined as in Section \ref{sec_prices}.\\

Using the same ideas as in \cite{gueant2017optimal} and with the same notations as above, we find that the markups that maximize our modified objective function are 
\begin{align}\label{optquotes01liqH}
 \delta^{0,1*}(t,z) = \bar \delta^{0,1} \left(z, \frac{\theta(t,Y_{t-}, \lambda^{0,1}_{t-}, \lambda^{1,0}_{t-}) -  \theta(t,Y_{t-}-z, \lambda^{0,1}_{t-}, \lambda^{1,0}_{t-}) }{z} \right)
 \end{align}
 and
\begin{align}\label{optquotes10liqH}
\delta^{1,0*}(t,z) = \bar \delta^{1,0} \left(z, \frac{\theta(t,Y_{t-}, \lambda^{0,1}_{t-}, \lambda^{1,0}_{t-}) -  \theta(t,Y_{t-}+z, \lambda^{0,1}_{t-}, \lambda^{1,0}_{t-}) }{z} \right).   
 \end{align}

\subsubsection{Curse of dimensionality}

The principal challenge with the above model is its dimensionality. In the basic geometric Brownian setting for prices, solving the optimal stochastic control problems \eqref{soc_liqH} necessitates a value function with time and three state variables. Incorporating the models from Section \ref{sec_prices} would increase the number of state variables to four or even five. Relying on such high-dimensional equations for a market making strategy is highly problematic. Moreover, the situation becomes even more critical as the quadratic Hamiltonian approximation technique, often utilized in Section \ref{sec_prices}, does not provide a means to reduce this dimensionality. This limitation arises because assuming the value function to be quadratic in $y$ does not result in linear terms in $y$ when computing finite differences in the Hamiltonian terms, due to simultaneous jumps in both the $y$ variable and the intensity variables.\\

While liquidity models based on Hawkes processes may offer a good statistical representation, they prove inadequate for our stochastic optimal control problems. The next section presents an alternative approach that effectively sidesteps the complications associated with Hawkes processes, thereby providing a more tractable solution for incorporating varying liquidity into our quoting strategies.

\subsection{Markov-modulated Poisson processes as a reasonable  alternative}

\subsubsection{Model, Hamilton-Jacobi-Bellman equation and optimal markups}

The main challenge with the liquidity model based on Hawkes processes is the simultaneity of trades and changes in intensities. To disentangle these two aspects and thus circumvent the curse of dimensionality, we can adopt a framework involving Markov-modulated Poisson processes, as proposed in \cite{bergault2023modeling}. This approach implies that the process $(\lambda^{0,1}_t, \lambda^{1,0}_t)_{t \in \mathbb R_+}$ is a continuous-time Markov chain, taking values in a finite set $\{\lambda_1, \ldots, \lambda_p\}^2$ -- where $p$ is small, typically $2$ or $3$, to represent low, high, and possibly medium levels of liquidity. The transitions between these states are governed by a rate matrix $Q \in M_{p^2}$.\footnote{In what follows, we shall order the states in lexicographic order, i.e. $$(\lambda_1, \lambda_1), \ldots, (\lambda_1, \lambda_p), \ldots, (\lambda_p, \lambda_1), \ldots, (\lambda_p, \lambda_p).$$}\\

Assuming that the price process is a geometric Brownian motion with drift, i.e. $dS_t = \mu S_t  dt + \sigma S_t dW_t$, $Y_t = (q_t^1 - q_0^1) S_t$ has the  dynamics 
$$dY_t = \mu Y_t dt + \sigma Y_t dW_t +  \int_{z \in \mathbb R_+^*}\!\!\!\!\!\!\!\!  z \left(J^{1,0}(dt,dz) -  J^{0,1}(dt,dz) \right)$$ and we see therefore that $(Y_t, \lambda_t^{0,1}, \lambda_t^{1,0})_{t \in \mathbb R_+}$ is Markovian as above.\\

The stochastic optimal control problems \eqref{sto_opt_cont} writes \eqref{soc_liqH} as above but the nature of the value function and the Hamilton-Jacobi-Bellman are different. For a given $\gamma$, the value function is a function $\theta:[0,T]\times \mathbb R \times \{\lambda_1, \ldots, \lambda_p\} \times \{\lambda_1, \ldots, \lambda_p\}\rightarrow \mathbb{R}$ and the associated Hamilton-Jacobi-Bellman equation is
\begin{equation}
\begin{cases}
 \!&0 = \partial_t \theta(t,y, \lambda_{j^{0,1}}, \lambda_{j^{1,0}}) + \mu y \left( 1 + \partial_y \theta(t,y, \lambda_{j^{0,1}}, \lambda_{j^{1,0}}) \right) - \frac{\gamma}{2} \sigma^2 y^2 + \frac 12 \sigma^2 y^2 \partial^2_{yy} \theta(t,y, \lambda_{j^{0,1}}, \lambda_{j^{1,0}})\\
\!& \qquad +\underset{1\le k^{0,1}, k^{1,0} \le p}{\sum}  Q_{(j^{0,1}-1) p+j^{1,0}, (k^{0,1}-1) p+k^{1,0}} \theta(t,y, \lambda_{k^{0,1}}, \lambda_{k^{1,0}})\\
\!& \qquad + \text{\scalebox{0.6}[1]{$\bigint$}}_{\!\!\mathbb{R}_{+}^{*}} \left(z\lambda_{j^{0,1}}\phi^{0,1}(z)H^{0,1} \left(z,\frac{\theta(t,y, \lambda_{j^{0,1}}, \lambda_{j^{1,0}}) -  \theta(t,y- z, \lambda_{j^{0,1}}, \lambda_{j^{1,0}}) }{z}\right)\right.\\
\!&\qquad\qquad\quad \left. + z\lambda_{j^{1,0}}\phi^{1,0}(z)H^{1,0} \left(z,\frac{\theta(t,y, \lambda_{j^{0,1}}, \lambda_{j^{1,0}}) -  \theta(t,y+z, \lambda_{j^{0,1}}, \lambda_{j^{1,0}}) }{z}\right) \right)m(dz),\\
\!&\theta(T,y, \lambda_{j^{0,1}}, \lambda_{j^{1,0}}) = 0,
\end{cases}
\label{eqn:HJBliqMMPP}
\end{equation} for all $\lambda_{j^{0,1}}, \lambda_{j^{1,0}} \in  \{\lambda_1, \ldots, \lambda_p\}$, where $H^{0,1}$ and $H^{1,0}$ are defined as in Section \ref{sec_prices}.\\

Using the same ideas as in \cite{gueant2017optimal} and with the same notations as above, we find that the markups that maximize our modified objective function are 
\begin{align}\label{optquotes01liqMMPP}
 \delta^{0,1*}(t,z) = \bar \delta^{0,1} \left(z, \frac{\theta(t,Y_{t-}, \lambda^{0,1}_{t-}, \lambda^{1,0}_{t-}) -  \theta(t,Y_{t-}-z, \lambda^{0,1}_{t-}, \lambda^{1,0}_{t-}) }{z} \right)
 \end{align}
 and
\begin{align}\label{optquotes10liqMMPP}
\delta^{1,0*}(t,z) = \bar \delta^{1,0} \left(z, \frac{\theta(t,Y_{t-}, \lambda^{0,1}_{t-}, \lambda^{1,0}_{t-}) -  \theta(t,Y_{t-}+z, \lambda^{0,1}_{t-}, \lambda^{1,0}_{t-}) }{z} \right).   
 \end{align}

\subsubsection{Discussion and numerical methods}

Under the above dynamics for the intensities, computing the optimal markups primarily involves solving the partial integro-differential equations \eqref{eqn:HJBliqMMPP} to derive the value function $\theta$, which depends on time and three state variables: the spread to Hodl $y$, and the two intensities $\lambda^{0,1}$ and $\lambda^{1,0}$. While this setup may appear similar to the liquidity model based on Hawkes processes, here the significant difference is that the two intensities take a very limited number of values, $p$ each. Notably, there are no partial derivative terms with respect to the intensities. Consequently, Eq. \eqref{eqn:HJBliqMMPP} should be regarded as a system of partial integro-differential equations in two dimensions: time and the spread to Hodl ($y$). This simplification effectively removes the issue of high dimensionality, and one could even consider extending the price model using those discussed in Section \ref{sec_prices}.\\

Once $\theta$ is computed using for instance operator splitting and an implicit scheme on a set of $p^2$ rectangular grids, one  can plug the current value of $y$ and an estimate\footnote{See \cite{bergault2023modeling} for estimation techniques.} of the current value of $(\lambda^{0,1}, \lambda^{1,0})$ in Eqs. \eqref{optquotes01liqH} and \eqref{optquotes10liqH} to get the optimal markups.\\

Interestingly, the quadratic Hamiltonian approximation technique works very well here and it leads to a system of ODEs when prices are log-normal (we would obtain PDEs with the price models of Section \ref{sec_prices}). Let us indeed define $\check \theta$ as the approximation of $\theta$ associated with the functions $$\check{H}^{0,1}: p \mapsto  \alpha^{0,1}_0 + \alpha^{0,1}_1 p + \frac 12 \alpha^{0,1}_2 p^2 \quad \textrm{and} \quad \check{H}^{1,0}: p \mapsto  \alpha^{1,0}_0 + \alpha^{1,0}_1 p + \frac 12 \alpha^{1,0}_2 p^2.$$
$\check \theta$ solves for all $\lambda_{j^{0,1}}, \lambda_{j^{1,0}} \in  \{\lambda_1, \ldots, \lambda_p\}$
\begin{equation}
\begin{cases}
 \!&0 = \partial_t \check \theta(t,y, \lambda_{j^{0,1}}, \lambda_{j^{1,0}}) + \mu y \left( 1 + \partial_y \check \theta(t,y, \lambda_{j^{0,1}}, \lambda_{j^{1,0}}) \right) - \frac{\gamma}{2} \sigma^2 y^2 + \frac 12 \sigma^2 y^2 \partial^2_{yy} \check \theta(t,y, \lambda_{j^{0,1}}, \lambda_{j^{1,0}})\\
\!& \qquad +\underset{1\le k^{0,1}, k^{1,0} \le p}{\sum}  Q_{(j^{0,1}-1) p+j^{1,0}, (k^{0,1}-1) p+k^{1,0}} \check \theta(t,y, \lambda_{k^{0,1}}, \lambda_{k^{1,0}})\\
\!& \qquad + \text{\scalebox{0.6}[1]{$\bigint$}}_{\!\!\mathbb{R}_{+}^{*}} \left(z \lambda_{j^{0,1}} \phi^{0,1}(z) \check H^{0,1} \left(z,\frac{\check \theta(t,y, \lambda_{j^{0,1}}, \lambda_{j^{1,0}}) -  \check \theta(t,y- z, \lambda_{j^{0,1}}, \lambda_{j^{1,0}}) }{z}\right)\right.\\
\!&\qquad\qquad\quad \left. + z\lambda_{j^{1,0}}\phi^{1,0}(z) \check H^{1,0} \left(z,\frac{\check \theta(t,y, \lambda_{j^{0,1}}, \lambda_{j^{1,0}}) -  \check \theta(t,y+z, \lambda_{j^{0,1}}, \lambda_{j^{1,0}}) }{z}\right) \right)m(dz),\\
\!&\check \theta(T,y, \lambda_{j^{0,1}}, \lambda_{j^{1,0}}) = 0.
\end{cases}
\label{eqn:quadHJBliqMMPP}
\end{equation}

Eq. \eqref{eqn:quadHJBliqMMPP} has a solution of the form $\check \theta(t,y, \lambda_{j^{0,1}}, \lambda_{j^{1,0}}) = - A(t, \lambda_{j^{0,1}}, \lambda_{j^{1,0}})y^2 -  B(t, \lambda_{j^{0,1}}, \lambda_{j^{1,0}})y - C(t, \lambda_{j^{0,1}}, \lambda_{j^{1,0}})$, where $(t,\lambda_{j^{0,1}}, \lambda_{j^{1,0}}) \mapsto A(t,\lambda_{j^{0,1}}, \lambda_{j^{1,0}}) \in \mathbb R$, $(t,\lambda_{j^{0,1}}, \lambda_{j^{1,0}}) \mapsto B(t,\lambda_{j^{0,1}}, \lambda_{j^{1,0}}) \in \mathbb R$ and $(t,\lambda_{j^{0,1}}, \lambda_{j^{1,0}}) \mapsto C(t,\lambda_{j^{0,1}}, \lambda_{j^{1,0}}) \in \mathbb R$ solve a system of ODEs. As the value of $C$ is irrelevant for what follows, we only report the equations for $A$ and $B$:
\begin{align}\label{ODEsysliqMMPP}
\begin{cases}
\partial_tA(t,\lambda_{j^{0,1}}, \lambda_{j^{1,0}}) &= 2A(t,\lambda_{j^{0,1}}, \lambda_{j^{1,0}})^2 \text{\scalebox{0.6}[1]{$\bigint$}}_{\!\!\mathbb{R}_{+}^{*}} z \left(\lambda_{j^{0,1}}\phi^{0,1}(z)\alpha^{0,1}_2(z) + \lambda_{j^{1,0}}\phi^{1,0}(z)\alpha^{1,0}_2(z) \right)m(dz) - \frac{\gamma}{2}\sigma^2\\
&\quad -\underset{1\le k^{0,1}, k^{1,0} \le p}{\sum}  Q_{(j^{0,1}-1) p+j^{1,0}, (k^{0,1}-1) p+k^{1,0}} A(t,\lambda_{k^{0,1}}, \lambda_{k^{1,0}}),\\ 
\partial_tB(t,\lambda_{j^{0,1}}, \lambda_{j^{1,0}})  &= -\underset{1\le k^{0,1}, k^{1,0} \le p}{\sum}  Q_{(j^{0,1}-1) p+j^{1,0}, (k^{0,1}-1) p+k^{1,0}} B(t,\lambda_{k^{0,1}}, \lambda_{k^{1,0}}) \\
&\qquad  + 2A(t,\lambda_{j^{0,1}}, \lambda_{j^{1,0}})\text{\scalebox{0.6}[1]{$\bigint$}}_{\!\!\mathbb{R}_{+}^{*}} z \left(\lambda_{j^{0,1}}\phi^{0,1}(z)\alpha^{0,1}_1(z) - \lambda_{j^{1,0}}\phi^{1,0}(z)\alpha^{1,0}_1(z) \right)m(dz)\\
&\qquad  - 2A(t,\lambda_{j^{0,1}}, \lambda_{j^{1,0}})^2\text{\scalebox{0.6}[1]{$\bigint$}}_{\!\!\mathbb{R}_{+}^{*}} z^2 \left(\lambda_{j^{0,1}}\phi^{0,1}(z)\alpha^{0,1}_2(z) - \lambda_{j^{1,0}}\phi^{1,0}(z)\alpha^{1,0}_2(z) \right)m(dz) \\
&\qquad + 2A(t,\lambda_{j^{0,1}}, \lambda_{j^{1,0}})B(t,\lambda_{j^{0,1}}, \lambda_{j^{1,0}})\\
& \qquad \quad \times\text{\scalebox{0.6}[1]{$\bigint$}}_{\!\!\mathbb{R}_{+}^{*}} z \left(\lambda_{j^{0,1}}\phi^{0,1}(z)\alpha^{0,1}_2(z) + \lambda_{j^{1,0}}\phi^{1,0}(z)\alpha^{1,0}_2(z) \right)m(dz)  ,\\
A(T,\lambda_{j^{0,1}}, \lambda_{j^{1,0}}) &= 0,\\
B(T,\lambda_{j^{0,1}}, \lambda_{j^{1,0}}) &= 0,
\end{cases}
\end{align}
for all $\lambda_{j^{0,1}}, \lambda_{j^{1,0}} \in  \{\lambda_1, \ldots, \lambda_p\}$.\\

The system of ODEs \eqref{ODEsysliqMMPP} can be solved straightforwardly. Once $A$ and $B$ are obtained, approximations of the optimal strategies can be computed by replacing $\theta$ by $\check \theta$ in Eqs. \eqref{optquotes01liqMMPP} and \eqref{optquotes10liqMMPP}. We thereby obtain 
\begin{align}\label{quadoptquotes01liqMMPP}
 \check \delta^{0,1*}(t,z) = \bar \delta^{0,1} \Big(z, A(t,\lambda^{0,1}_{t-}, \lambda^{1,0}_{t-}) \left(z-2Y_{t-} \right) -B(t,\lambda^{0,1}_{t-}, \lambda^{1,0}_{t-}) \Big)
 \end{align}
 and
\begin{align}\label{quadoptquotes10liqMMPP}
\check \delta^{1,0*}(t,z) = \bar \delta^{1,0} \Big(z, A(t,\lambda^{0,1}_{t-}, \lambda^{1,0}_{t-}) \left(z+2Y_{t-} \right) +B(t,\lambda^{0,1}_{t-}, \lambda^{1,0}_{t-}) \Big).   
 \end{align}

\section*{Conclusion}
\label{concl}

In this paper, we introduced a comprehensive suite of models tailored for price-aware automated market makers designed to optimize their quoting strategies. Our exploration spanned a variety of advanced price dynamics, including stochastic volatility, price jumps, and microstructural models based on Hawkes processes. However, the adoption of these sophisticated models brings challenges, particularly in terms of higher computational costs, which must be considered as they can restrict practical application. Additionally, we addressed issues related to liquidity modeling, specifically the challenges associated with the use of Hawkes processes and proposed a viable alternative through Markov-modulated Poisson processes.\\

The selection of an appropriate model, whether for prices or liquidity dynamics, depends on the specific cryptocurrency pair involved and ultimately lies in the hands of the developers of these AMMs. They must balance the benefits of detailed, complex modeling against computational efficiency and the operational constraints of their systems. The analyses presented in this paper should illuminate the path forward for the industry, guiding the integration of these advanced models into practical, real-world applications.

\section*{Acknowledgement}

The authors wish to express their gratitude to the management of Swaap Labs for their support throughout the research process. They presented their work at several conferences and seminars, benefiting greatly from the audiences' insightful remarks. Special thanks are also extended to \'{A}lvaro Cartea and Fayçal Drissi for their valuable comments on a previous version of the manuscript.

\section*{Statement on funding}

The research carried out for this paper benefited from the support of the Research Program ‘‘Decentralized Finance and Automated Market Makers’’, a Research Program under the aegis of Institut Europlace de Finance, in partnership with Swaap Labs. David Bouba and Julien Guilbert are both members of the Swaap Labs team which is an actor of the DeFi industry.

\section*{Statement on data availability}

No data involved in this study.

\bibliographystyle{plain}

\end{document}